\documentclass[journal]{IEEEtran}
\IEEEoverridecommandlockouts
\usepackage[subrefformat=parens]{subcaption}
\usepackage{amsmath,graphicx}
\usepackage{amsmath,amssymb,amsfonts}
\usepackage{cite, soul}
\usepackage{url}
\usepackage{bm}
\usepackage{comment}

\usepackage[usenames,dvipsnames]{color}

\newtheorem{remark}{Remark}

\usepackage[linesnumbered,vlined,ruled]{algorithm2e}
\usepackage{algpseudocode}

\usepackage{tabularx}
\usepackage{makecell}

\title{LISAC: Learned Coded Waveform Design \\ for ISAC with OFDM}
\author{Chenghong Bian, \IEEEmembership{Student Member,~IEEE},
Yumeng Zhang, \IEEEmembership{Member,~IEEE}, Meng Hua, \IEEEmembership{Member,~IEEE}, \\
Kaitao Meng, \IEEEmembership{Member,~IEEE},
Deniz G{\"u}nd{\"u}z, \IEEEmembership{Fellow,~IEEE}
\thanks{C. Bian, M.Hua and D. G{\"u}nd{\"u}z are with the Department of Electrical and Electronic Engineering, Imperial College London (E-mails: \{c.bian22, m.hua, d.gunduz\}@imperial.ac.uk). {Y. Zhang is with the Department of Electrical and Computer Engineering, Hong Kong University of Science and Technology, Hong Kong. (E-mail: eeyzhang@ust.hk)}. K. Meng is with the Department of Electrical and Electronic Engineering, University of Manchester, Manchester, UK (E-mail: kaitao.meng@manchester.ac.uk)
}
\thanks{This work received funding from the UKRI for the projects AI-R (ERC Consolidator Grant, EP/X030806/1) and the SNS JU project 6G-GOALS under the EU’s Horizon program Grant Agreement No. 101139232.}
\thanks{This paper was presented in part at the IEEE Wireless Communications and Networking Conference, 2025 \cite{lisac_conf}.}
}

\begin{document}

\maketitle
\begin{abstract}
We propose deep learning based coded waveform design for integrated sensing and communication (ISAC) with orthogonal frequency-division multiplexing (OFDM).  
Our goal is to design a coded waveform capable of delivering accurate target parameter estimation while maintaining high communication quality measured in terms of bit error rate (BER). 
In the proposed learned coded waveform for ISAC (LISAC), the pilot and data encoding functions at the encoder are parameterized by recurrent neural networks (RNNs) and  are trained jointly in an end-to-end fashion.  
The communication receiver estimates the channel and performs residual-assisted minimum mean square error (MMSE) channel equalization, where a neural network is introduced to calibrate the coarse estimate produced by the standard MMSE channel equalizer. Then, an RNN-based channel decoder is employed to decode the information bits using the equalized signal. 
Two different sensing loss functions are considered, one calculates the mean square error (MSE) between the original and the estimated sensing parameters, while the other calculates the Cramér-Rao lower bound (CRLB).
The LISAC modules are optimized using a weighted combination of communication and sensing losses and different trade-off points between the sensing and communication performances are achieved by adjusting the weights. 
Simulation results show that the proposed LISAC waveform achieves a better trade-off curve compared to existing alternatives for both AWGN and multi-path fading scenarios. Ablation studies are carried out to demonstrate the gain brought by each design component for a better understanding of the proposed scheme.
\end{abstract}
%


\section{Introduction}
\label{sec:intro}

\begin{figure}[t]
     \centering
         \includegraphics[width=\columnwidth]{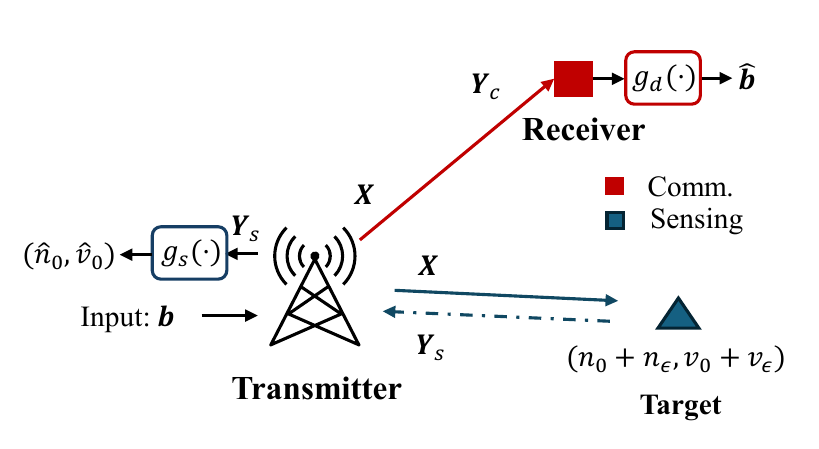}  
  \caption{The considered ISAC system model where the transmitter is sensing a target while communicating with a receiver.  The reflection of the signal carrying the information bits is used by the transmitter to estimate the target parameters.}
  \label{fig:fig_system}
\end{figure}

Integrated sensing and communications  is attracting increasing attention \cite{isac1, hua2024integrated,hua2023joint,tit_isac} due to its great potential in mitigating the spectrum shortage and reducing the hardware costs, paving the way for new applications and services in the upcoming 6G networks \cite{meng2024cooperative}. 
ISAC offers a unified framework which enables the transmission of information bits while simultaneously estimating the parameters of interest, such as distance, velocity, and angle-of-arrival.
Ideal ISAC waveform should balance the communication rate with the detection/estimation performance for sensing.
Designing an effective ISAC waveform is challenging due to the conflicting requirements of sensing and communications. A foundational study on the trade-off between these two functions from an information-theoretical perspective is presented in \cite{tit_isac}, {where the authors show that the sensing-optimal input, based on the CRLB, features a deterministic signal,} while the communication-optimal input follows a random Gaussian distribution, highlighting a so-called ``deterministic-random trade-off'' between sensing and communications. 
Rather than designing a new ISAC waveform from scratch, a more pragmatic approach that has been widely employed in the literature is to employ OFDM waveform for sensing, which has been shown to provide promising estimation precision in radar applications \cite{ofdm_radar_overview, ofdm_radar_ml, bian2024sparseregressioncodesintegrated}. 
A potential solution to the deterministic-random trade-off in OFDM-based ISAC is to use PSK modulation \cite{ofdm_radar}, but this limits the communication rate. The superiority of OFDM among communication-centric ISAC waveforms is shown in \cite{liu2024ofdmachieveslowestranging} for both PSK and QAM symbols by studying the ranging performance assuming random information symbols.
{The authors in \cite{isac_waveform_design} propose outlier mean square error (MSE) and outlier probability (OP) as the sensing metrics accounting for OFDM's resolution limits}. 
However, \cite{isac_waveform_design} optimizes the OP metric under an asymptotic channel capacity constraint. 
Instead of focusing on a uniform distribution, probabilistic constellation shaping is considered in \cite{du2023reshapingisactradeoffofdm} to balance sensing performance with channel capacity.


While all the above works evaluate the communication performance through the asymptotic channel capacity achieved through independent and identically distributed (i.i.d.) channel symbols, here we are interested in the practical channel coding performance. It is known that without channel codes, the BER performance of a communication system would remain in a high level even under a high SNR. There are few prior works that study the feasibility of applying channel codes to  ISAC systems. The authors in \cite{isac_svc} explore the potential of sparse vector codes (SVCs) as a viable ISAC waveform, showing that they surpass convolutional codes in both error correction and sensing capabilities. In \cite{isac_code}, the authors demonstrate that low-density parity-check (LDPC) coded signals does not degrade the sensing performance with respect to (w.r.t.) uncoded signals. Channel codes are also shown to be effective in the passive sensing scenario. In particular, it is shown in \cite{bian2024sparseregressioncodesintegrated} that by adopting cyclic redundancy check assisted channel codes with an iterative parameter sensing and channel decoding algorithm, superior packet error rate and sensing performances can be obtained over uncoded baselines \cite{two_stage_ipsac, isac_learn}.

Recently, deep neural networks (DNNs) have been successfully applied to the design of channel codes \cite{decode_linear_code, choukroun2022error, kim2018communication, LEARN_code, turboae, learn_isac, feedback_code}. The pioneer work \cite{decode_linear_code} employs learnable neural network parameters to reinforce the belief propagation (BP) algorithm for the decoding of BCH and LDPC codes. More advanced neural network architectures, e.g., RNNs and transformers have also been utilized to facilitate the decoding of convolutional \cite{kim2018communication} and linear block codes \cite{choukroun2022error}, achieving similar or even superior BER performance compared with their corresponding state-of-the-art decoding algorithms.  
Researchers have also used neural networks to invent completely new codes. In \cite{LEARN_code}, the authors parameterize the encoder and decoder using 1D convolutional neural networks. This approach is extended in \cite{turboae}, where iterative neural decoding is employed to achieve comparable BER performance w.r.t. conventional turbo codes. Transformer architecture is employed in \cite{feedback_code} to achieve extremely low BER even at very short blocklengths in the presence of feedback. Despite some performance improvements reported in these works, they may not be sufficient for the deployment of DNN-aided channel codes in practice considering their increased computational and memory requirements. However, as we show in this paper, gains can be much more significant when sensing performance is also considered. 
There are only a few works which employ deep learning to design codes for ISAC systems. A preliminary study is carried out in \cite{learn_isac}, where a simple DNN is employed to design a short block length code ($\sim 6$ bits) for non-coherent communication and sensing. Our prior work \cite{lisac_conf} designed a learned coded waveform for ISAC systems under the assumption of an AWGN communication channel. However, in practical scenarios, the communication link is subject to multipath fading, which necessitates more sophisticated processing techniques, such as the design of pilot signals and the corresponding channel estimation.

In this paper, we develop a deep learning based coded waveform, called LISAC, to achieve both satisfactory sensing performance and error correction ability based on OFDM. As illustrated in Fig. \ref{fig:fig_system}, we are interested in reliably transmitting the information bits to the receiver and accurately estimating the delay and Doppler parameters of the target.
LISAC transmitter comprises of separate pilot and data encoding functions. 
Inspired by the great success of learned  pilot signal design to achieve superior channel estimation performance as in \cite{learn_a_pilot, tvt_deep_pilot}, we consider a learning-based approach where the pilot symbols are produced by RNNs, which are jointly optimized with the data symbols. This enables not only flexible power allocation between the pilot and data symbols, but also achieves both superior sensing performance and improved channel estimation quality.
At the receiver, the Minimum Mean Square Error (MMSE) channel estimation is performed to obtain  channel state information (CSI), followed by a neural network aided MMSE channel equalization module. The former produces a coarse estimate of the transmitted symbols, while the latter, parameterized by a RNN, generates a more refined estimate. Finally, a likelihood estimate for each  information bit is generated by the channel decoder.
The loss functions for the training of the LISAC model are carefully chosen. In particular, two sensing loss functions are considered, where the first one calculates the outlier MSE, i.e., the $l_2$ distance between the true delay and Doppler parameters and the estimated ones, using maximum likelihood (ML) principle \cite{isac_waveform_design}. We also consider the CRLB as an alternative sensing loss and adopts the well-known cross entropy loss to optimize the BER performance.
The encoder and decoder are trained alternatively using different weighted combinations of the communication and sensing losses to achieve different points on the sensing and communication trade-off plane.  Numerical experiments show that the learned code spreads the transmitted signals across the I/Q plane when communication performance is the priority, while it converges to a mixture between PSK and amplitude modulation when the sensing performance becomes important. Our results show that the proposed learned coded-modulation scheme achieves a better sensing and communication trade-off compared with existing methods evaluated under both the AWGN and multi-path fading channels. 

The contributions of this paper are summarized as follows:
\begin{itemize}
    \item We propose LISAC, the first deep learning-based waveform design framework for ISAC based on OFDM.   Transceivers of LISAC are parameterized by RNNs and trained using a weighted loss function combining sensing and communication objectives to achieve different sensing and communication trade-offs. 

    \item We present dedicated LISAC transceiver design to explore learning-based channel codes under the OFDM setup. The proposed OFDM symbol mapping strategy alleviates possible deep fades introduced by the frequency-selective channel in the consecutive OFDM data symbols, and the proposed residual-assisted MMSE channel equalization method produces more accurate estimate of the transmitted signal, significantly improving the  system  performance.

    \item A pilot encoding function, parameterized also by a RNN, is employed to learn the pilots jointly with the data symbols in an end-to-end fashion. Compared with conventional schemes with unitary pilot symbols,  learned pilot encoding  enables a more flexible waveform design, and thus improves both the sensing and communication performances.

    \item Extensive numerical experiments are carried out to show the effectiveness of the proposed LISAC model compared with existing schemes in terms of both BER and sensing performances under AWGN and multi-path fading channels. Ablation studies are carried out to evaluate the gains from each neural network module for a comprehensive understanding.  
    
\end{itemize}

{\it Notations:} Throughout the paper, normal-face letters (e.g., $x$) represent scalars, while uppercase letters (e.g., $X$) represent random variables. Matrices and vectors are denoted by bold {upper} and {lower} case letters (e.g., $\bm{X}$ and $\bm{x}$), respectively.   
$\Re(x)$ ($\Im(x)$) denotes the real (imaginary) part of {a complex variable} $x$.
Transpose and Hermitian operators are denoted by $(\cdot)^\top$, $(\cdot)^\dagger$, respectively. $\text{vec}(\cdot)$ represents the vectorization operation and $\text{diag}(\bm{x})$ outputs a diagonal matrix with $\bm{x}$ as its diagonal elements. ${x}[i]$ denotes the $i$-th element of the vector, $\bm{x}$, while $\bm{1}_{K\times M}$ represents a $K\times M$ matrix whose elements are all ones.
{Finally, $\|\bm{S}\|_F$  denotes the Frobneous norm of the matrix $\bm{S}$.}

\begin{figure*}
     \centering

         \includegraphics[width=2\columnwidth]{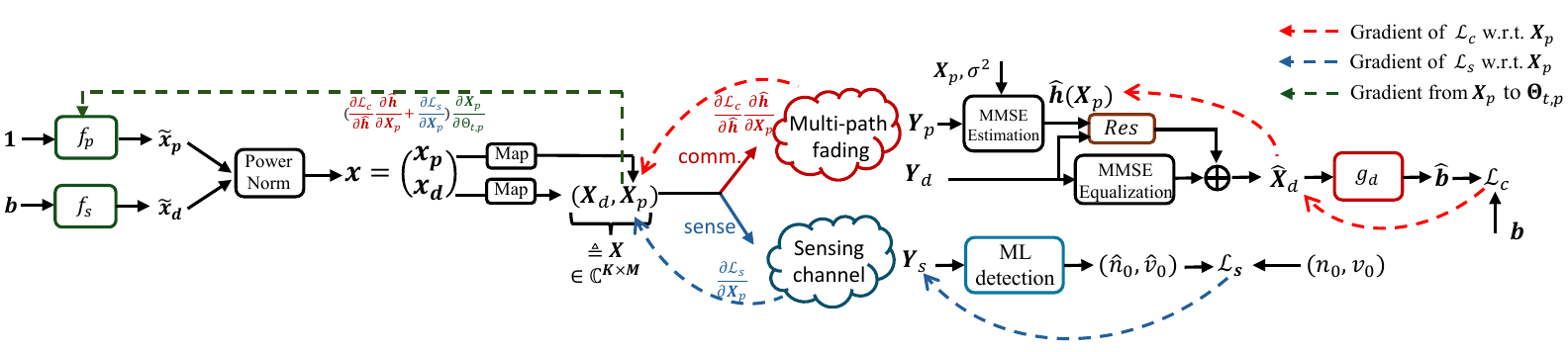}

  \caption{The proposed LISAC framework. LISAC encoder generates pilot and data symbols, $\bm{X}_p$ and $\bm{X}_d$, via the encoding functions, $f_p(\cdot)$ and {$f_s(\cdot)$}, respectively, followed by power normalization and OFDM symbol mapping modules. Receiver's goal is to recover the bit sequence with the help of the pilot sequence to estimate the CSI. Transmitter estimates delay and Doppler parameters, $(\hat{n}_0, \hat{v}_0)$, of the target using the reflection of the transmitted waveform. We explicitly illustrate the gradients w.r.t. the pilot signal $\bm{X}_p$ with dashed lines in the figure, which are utilized to update the parameters, $\Theta_{t,p}$, of the RNN used as the pilot encoding function, $f_p(\cdot)$.}
\label{fig:system_model}
\end{figure*}

\section{Problem Formulation}\label{sec:problem}
In this section, we introduce the ISAC scenario, where both pilot and coded data symbols are utilized to perform reliable communication and target sensing simultaneously.

\subsection{Communication model}\label{sec:comm_model}
We consider a communication system using OFDM where each OFDM frame is comprised of  pilot and data symbols, which are denoted as $\bm{X}_p$ and $\bm{X}_d$, respectively. For ease of analysis, the pilot signal, $\bm{X}_p \in \mathbb{C}^{K\times M_p}$, is assumed to take up the first $M_p$ OFDM symbols, across $K$ subcarriers. The  data symbols, $\bm{X}_d \in \mathbb{C}^{K\times M_d}$, are allocated the remaining $M_d$ symbols, and generated  via an encoding function, $f_s(\cdot)$, from the information bit sequence, $\bm{b} \in \{0, 1\}^{N_b}$, i.e.,:
\begin{equation}
\bm{X}_d = f_s(\bm{b}).
\label{equ:fs}
\end{equation}
The overall transmitted signal is denoted by $\bm{X} \triangleq [\bm{X}_p, \bm{X}_d]$ occupying a total number of $M \triangleq M_p + M_d$ OFDM symbols.
A power constraint is imposed on the entire OFDM frame, $\bm{X}$, as:
\begin{equation}
\|\bm{X}\|_F^2 \leq KM.
\label{equ:pn}
\end{equation}
OFDM subcarriers belonging to the $m$-th OFDM symbol, $\bm{X}[:, m]$, are transformed to the time-domain signals, denoted as $\bm{x}[:, m]$ via IFFT, and is concatenated with a length-$K_G$ cyclic prefix (CP) for transmission. 

We consider an $L$-tap multi-path fading channel, where the $\ell$-th path is characterized by its complex radar cross section (RCS), $\alpha_\ell$, and delay, $\tau_\ell$, for $\ell \in [1, L]$. We assume the maximum delay does not exceed the length of the CP, i.e., $\tau_\ell \in [0, K_G)$. Moreover, the RCS parameters, $\alpha_\ell$, are identically independently distributed (i.i.d.), and follow complex Gaussian distribution, i.e., $\alpha_\ell \sim \mathcal{CN}(0, \frac{1}{L})$. The CSI in the frequency domain is obtained by applying FFT to the time domain response and is denoted by $\bm{h} \in \mathbb{C}^{K}$. We assume a block static channel where $\bm{h}$ remains the same over the entire OFDM frame, $\bm{X}$.

After removing the CP and applying FFT, the $m$-th, received OFDM symbol, $m \in [1, M]$ can be expressed as:
\begin{equation}
\bm{Y}_c[:, m] = \underbrace{\bm{h} \odot \bm{X}[:, m]}_{\triangleq\bm{S}[:, m]} + \bm{W}_c[:, m],
\label{equ:comm_channel}
\end{equation}
where $\odot$ represents element-wise multiplication, $\bm{S}$ denotes the noiseless received signal, and $\bm{W}_c \in \mathbb{C}^{K\times M}$ is the additive white Gaussian noise (AWGN) in the communication link, satisfying ${W}_c[i, j] \sim \mathcal{CN}(0, \sigma_c^2)$. The received signal can be partitioned into the received pilot and data symbols, i.e., $\bm{Y}_c = [\bm{Y}_p, \bm{Y}_d]$. The signal-to-noise-ratio (SNR) for the communication link is 
\begin{equation}
    \mathrm{SNR}_c \triangleq 
\frac{\mathbb{E}(\|\bm{S}\|_F^2)}{\mathbb{E}(\|\bm{W}_c\|_F^2)} = 1/\sigma_c^2,
\end{equation}
which is due to the fact that ${\mathbb{E}(\|\bm{S}\|_F^2)} = KM$ given $\alpha_\ell \sim \mathcal{CN}(0, \frac{1}{L}), \ell \in [1, L]$.

At the receiver, after receiving the noisy signal, $\bm{Y}_c$, the standard MMSE channel estimation algorithm is performed to produce the estimated CSI, denoted as $\hat{\bm{h}}$. {We first consider $\bm{X}_p = \bm{1}_{K\times M_p}$ as pilot signal, and the corresponding MMSE channel estimate can be expressed as:}\footnote{The MMSE channel estimate obtained by non-unitary pilot symbols, $\bm{X}_p$, is illustrated in \eqref{eq:est_csi_plt}.}
\begin{equation}
    \hat{\bm{h}} = \frac{\sum_{i=1}^{M_p} \bm{Y}_p[:, i]}{{M_p+ \sigma_c^2}}.
    \label{eq:est_csi}
\end{equation}
Then, we equalize the received OFDM data symbols, $\bm{Y}_d$, using the estimated $\hat{\bm{h}}$ as follows:
\begin{equation}
    \hat{{X}}_d[k, m] =  \hat{{h}}_k {Y}_d[k, m] / (|\hat{{h}}_k|^2 + \sigma_c^2),
    \label{eq:equ_xki}
\end{equation}
where {$k \in [1, K], m \in [1, M_d]$} and $\hat{h}_k$ denotes the $k$-th element of $\hat{\bm{h}}$.
Finally, by feeding the equalized signal, $\bm{\hat{X}}_d$, into the channel decoder, we obtain the bit sequence, $\bm{\hat{b}}$ as:
\begin{equation}
\bm{\hat{b}}= g_d(\bm{\hat{X}}_d),
\label{equ:decode_bitseq}
\end{equation}
where $g_d(\cdot)$ denotes the decoding function, which first transforms $\bm{\hat{X}}_d$ into a vector of length-$KM_d$ and then demodulates and decodes the vector. The BER is utilized as the communication performance metric, defined as:
\begin{equation}
\text{BER} \triangleq \frac{1}{N_b} \mathbb{E} \left[ \sum_{i=1}^{N_b} \bm{1}(\hat{{b}}_i \neq {{b}}_i) \right],
\label{equ:ber_define}
\end{equation}
where $\bm{1}(\cdot)$ represents the indicator function.

We will also consider a simplified scenario with a single line-of-sight (LoS) path with a unitary channel gain \cite{lisac_conf}. In this case, the multi-path fading channel depicted in \eqref{equ:comm_channel} degrades to an AWGN channel, i.e., $\bm{h} = \bm{1}_K$. We will evaluate both channels in the simulation section.

\subsection{Sensing model}
We then describe the proposed sensing model, assuming a single target at a distance $R$ from the transmitter moving at a speed $v$. As a result, the received echo signal is delayed by $\tau = 2R/c$ with a Doppler shift of $f_D \triangleq f_c (2v/c)$, where $f_c$ is the carrier frequency. The bandwidth of the OFDM symbol, $B$, corresponds to a delay resolution of  $\Delta\tau = 1/B$. The duration of the OFDM symbol is $T = (K + K_G)/B$ leading to a Doppler resolution of $\Delta f_D =  \frac{B}{(K + K_G)M}$. The delay and Doppler of the target, $\tau$ and $f_D$, can be represented using $\Delta \tau$ and $\Delta f_D$, respectively:
\begin{align}
\tau &= \underbrace{(n_0 + n_\epsilon)}_{n^*} \Delta \tau, \notag \\
f_D &= \underbrace{(v_0 + v_\epsilon)}_{v^*} \Delta f_D,
\label{equ:delay_doppler}
\end{align}
where $(n^*, v^*)$ denote the normalized delay and Doppler parameters, and $(n_0, v_0)$ are their integer parts, respectively.  In this paper, we focus on the integer parts of the parameter estimates as the fractional delay and Doppler, $n_\epsilon \in (0, 1)$ and $v_\epsilon \in (-1/2, 1/2)$, are often out of the scope of estimation due to the radar resolution issues. Moreover, in practical implementations, the time-efficient 2D-FFT algorithm is adopted whose outputs are integer delay and Doppler bins. However, the effects of the fractional parts on the estimation of the parameters of interest, $(n_0, v_0)$, need to be taken into account. 

After removing the CP, the time-domain signal at the {$n$-th time slot} of the $m$-th OFDM symbol can be expressed as \cite[Eqn 5]{isac_waveform_design}:
\begin{equation}
{{y}_s[n, m] \approx a {x}[n-n_0, m]e^{j2\pi\frac{(v_0 +v_\epsilon)m}{M}} + {w}_s[n, m],}
\label{equ:td_y}
\end{equation}
where $a$ is the complex gain of the radar channel assumed to remain constant during the OFDM block/frame, while ${x}[n-n_0,m]$ is a circular shifted version of the transmitted signal ${x}[n,m]$ by $n_0$ due to the delay\footnote{The fractional term of the delay, $n_\epsilon$ is omitted because we assume perfect time synchronization at the receiver that eliminates the effect of $n_\epsilon$. This is reasonable since sensing is carried out by the transmitter itself.}. The  $e^{j2\pi\frac{(v_0 +v_\epsilon)m}{M}}$ term represents the phase shift of the $m$-th OFDM symbol due to the Doppler effect. Finally, ${w}_s[n, m] \in \mathcal{CN}(0, \sigma^2_s)$ denotes the AWGN term, and the SNR of the sensing link is defined as:
\begin{equation}
    \mathrm{SNR}_s \triangleq A^2/\sigma_s^2,
    \label{eq:sense_snr}
\end{equation}
where $A \triangleq |a|$ denotes the amplitude for the complex gain.

We then convert the time-domain signal in \eqref{equ:td_y} to the frequency domain via $K$-point FFT:
\begin{equation}
{Y}_s[k, m] = a{X}[k, m]e^{j2\pi\frac{(v_0 +v_\epsilon)m}{M}}e^{-j2\pi\frac{n_0k}{K}} + {W}_s[k, m],
\label{equ:sense_y}
\end{equation}
where ${W}_s[k,m] \sim \mathcal{CN}(0, \sigma^2_s)$ is the noise in the frequency domain.

\begin{figure*}
     \centering
     \begin{subfigure}{0.66\columnwidth}
         \centering
         \includegraphics[width=\columnwidth]{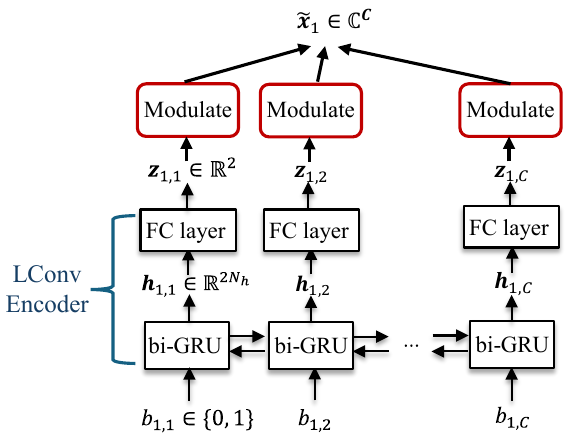}
         \caption{Data encoding function, $f_p(\cdot)$}
     \end{subfigure}
     \begin{subfigure}{0.66\columnwidth}
         \centering
         \includegraphics[width=0.8\columnwidth]{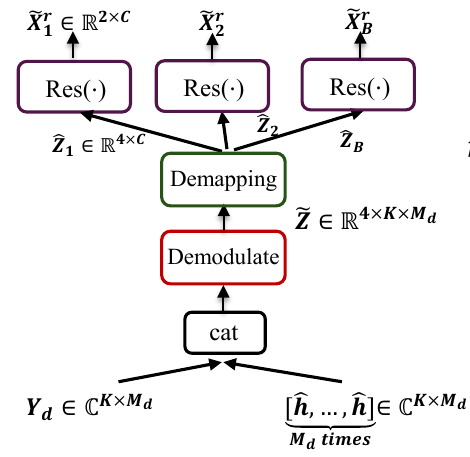}
         \caption{ResidualNet, $\text{Res}(\cdot)$}
     \end{subfigure}
     \begin{subfigure}{0.66\columnwidth}
         \centering
         \includegraphics[width=\columnwidth]{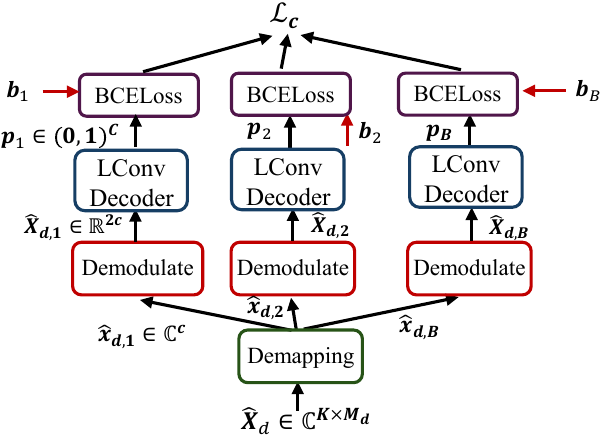}
         \caption{Data decoding function, $g_d(\cdot)$}
     \end{subfigure}
  \caption{Illustration of the neural network architectures of the data encoding function, $f_p(\cdot)$ at the transmitter, the ResidualNet, $\text{Res}(\cdot)$ and data decoding function, $g_d(\cdot)$ at the receiver. (a). The data encoding function is comprised of LConv module which takes the bit sequence, $\bm{b}_i, i\in [1, B]$ as input followed by a modulation block to generate the complex-valued codeword, $\tilde{\bm{x}}_i \in \mathbb{C}^C$. (b) The ResidualNet takes the received data symbols and the estimated CSI as input and outputs the residual signal to calibrate the MMSE equalization output.  (c). The LISAC decoder decodes the information bits from the equalized signal, $\hat{\bm{X}}_d$. The BCE loss is adopted as a loss function for the communication part.}
\label{fig:jsc_framework}
\end{figure*}

\subsubsection{Maximum likelihood (ML) estimation}
We then elaborate on the ML estimator of the target parameters of interest, i.e., $(n_0, v_0)$. The log-likelihood function can be expressed as:
\begin{align}
L(\bm{Y}_s|n_0, v_0, v_\epsilon, a) &\propto -\sum_{k=0}^{K-1}\sum_{m=0}^{M-1} \Big| {Y}_s[k,m] - \notag \\ &{X}[k, m]e^{j2\pi\frac{(v_0 +v_\epsilon)m}{M}}e^{-j2\pi\frac{n_0k}{K}}\Big|^2.
\label{equ:log_ll}
\end{align}
We note that the ML estimator of $a$ is independent of $n_0$ and $v_0$, hence is omitted here \cite{kay1993estimation}.
By expanding the right-hand-side (RHS) of \eqref{equ:log_ll}, the ML estimator can be expressed as:
\begin{align}
    (\hat{n}_0, \hat{v}_0) &= \arg \max_{(n, v)} \Big|\underbrace{\sum_{m,k} p_{k,m}e^{j2\pi\frac{nk}{K}}e^{-j2\pi\frac{(v + v_\epsilon)m}{M}}}_{\triangleq r_{n,v, v_{\epsilon}}} + {\widetilde{W}}_s[n, v] \Big| \notag \\
    & = \arg \max_{(n, v)} |\gamma_{n, v, v_\epsilon}|,
    \label{equ:ml_detect}
\end{align}
where $p_{k,m} \triangleq |{X}[k,m]|^2$ is the power of the transmit signal for the $k$-th subcarrier at the $m$-th OFDM symbol. We further note that $\gamma_{n, v, v_\epsilon}$ denotes the cross correlation between the received and the transmitted signals satisfying $\gamma_{n, v, v_\epsilon} \triangleq r_{n, v, v_\epsilon} + {\widetilde{W}}_s[n, v]$ where $r_{n, v, v_\epsilon}$ is the discrete ambiguity function while $\widetilde{{W}}_s[n,v], n\in [0, ~K_G-1], v\in [-M/2, ~ M/2-1]$ corresponds to the noise term defined as:
\begin{equation}
    {\widetilde{W}}_s[n,v] \triangleq  A^{-1}\sum_{m,k} {W}_s[k,m]{X}^*[k,m] e^{-j2\pi\frac{nk}{K}}e^{j2\pi\frac{vm}{M}}.
    \label{equ:noise_w}
\end{equation}
It can be noticed from \eqref{equ:ml_detect} that the ML estimator performance is determined by the power of the waveform, i.e., $p_{k,m}$. 

{
\subsubsection{Outlier MSE}
Due to the noise and the fractional Doppler, $v_\epsilon$, the ML detector output, $(\hat{n}_0, \hat{v}_0)$ might deviate from the ground truth, $(n_0, v_0)$. As in \cite{isac_waveform_design} and \cite{outlier_mse}, we adopt the outlier MSE, denoted as $E_o$, to quantify the expected $l_2$-distance between $(\hat{n}_0, \hat{v}_0)$ and $(n_0, v_0)$.
Without loss of generality, we assume $(n_0, v_0) = (0,0)$ and $E_o$ is expressed as:
\begin{equation}
    E_o \triangleq \mathbb{E}_{v_\epsilon} (\hat{n}_0^2 + \hat{v}_0^2),
    \label{equ:outlier_mse}
\end{equation}
}
where $v_\epsilon \sim \mathcal{U}(-\frac{1}{2}, \frac{1}{2})$ follows a uniform distribution.

\section{The proposed LISAC framework}\label{sec:LISAC}
Although \cite{isac_waveform_design} provides an insightful solution to the waveform design problem, i.e., $p_{k,m}$, using conventional optimization methods, the authors do not take the error correction ability of the optimized waveform into account. It is difficult to incorporate the BER performance into the conventional optimization frameworks; therefore, we employ a learning-based  approach to overcome this difficulty. The proposed LISAC system learns the OFDM pilots and the coded data symbols simultaneously for superior sensing and communication performances.

\subsection{LISAC Transmitter}
As shown in the LHS of Fig. \ref{fig:system_model}, in this subsection, we discuss the encoding functions, $f_p(\cdot)$ and $f_s(\cdot)$, of the LISAC encoder to generate the pilot and data OFDM symbols, $\bm{X}_p$ and $\bm{X}_d$, respectively. Both functions are parameterized by neural networks, denoted by $\bm{\Theta}_t \triangleq \{\bm{\Theta}_{t,p}, \bm{\Theta}_{t,s}\}$.

\subsubsection{Data symbol generation}
The input bit sequence, $\bm{b} \in \{0, 1\}^{N_b}$, is first partitioned into $B$ blocks, i.e., $\bm{b} = [\bm{b}_1, \bm{b}_2, \ldots, \bm{b}_B]$, where each block is of $C$ bits, i.e., $\bm{b}_b \in \{0, 1\}^C$, satisfying $BC = N_b$.
All the $B$ blocks are processed by the data encoding function, $f_s(\cdot)$, simultaneously. 

{As shown in Fig. \ref{fig:jsc_framework}(a)}, $\bm{b}_1$ is first fed to the encoder neural network, called LConv, to obtain a complex-valued codeword, $\tilde{\bm{x}}_1 \in \mathbb{C}^{C}$.
In particular, for each of the $c$-th time step,  the bidirectional-GRU (bi-GRU) block of the LConv encoder generates two hidden vectors, $\bm{h}_{1, c}^f, \bm{h}_{1, c}^b \in \mathbb{R}^{N_h}$ corresponding to the forward and backward directions, respectively where $N_h$ denotes the hidden dimension of the bi-GRU block.  For the forward direction, the neural network module, denoted as $\textit{bi-GRU}^f(\cdot)$ takes the $c$-th bit, ${b}_{1, c}$ and the forward hidden vector, $\bm{h}_{1, c-1}^f$ from the previous time step as input and outputs the hidden vector, $\bm{h}^f_{1, c}$ for the current time step:
\begin{equation}
    \bm{h}^f_{1, c} = \textit{bi-GRU}^f({b}_{1, c}, \bm{h}^f_{1, c-1}).
    \label{equ:bi-gru_f}
\end{equation}
The backward direction proceeds similarly:
\begin{equation}
    \bm{h}^b_{1, c} = \textit{bi-GRU}^b({b}_{1, c}, \bm{h}^b_{1, c+1}).
    \label{equ:bi-gru_b}
\end{equation}
We initialize $\bm{h}^f_{1, 0} = \bm{h}^b_{1, C+1} = \bm{0}_{N_h}$ to guarantee \eqref{equ:bi-gru_f} and \eqref{equ:bi-gru_b} hold for each time step $c \in [1, C]$.
By combining the hidden vectors corresponding to the forward and backward directions at all the $C$ time steps, we obtain the overall bi-GRU output as:
\begin{align}
    \{\bm{h}_{1, c}\}_{c\in [1, C]} &= \textit{bi-GRU}(\{{b}_{1, c}\}_{c\in [1, C]}), \\
    \text{where} & \quad \bm{h}_{1, c} \triangleq [(\bm{h}^f_{1, c})^\top, (\bm{h}^b_{1, c})^\top]^\top \notag.
    \label{equ:bi-gru}
\end{align}
Then, a fully connected (FC) layer is applied to each of the latent vectors, $\bm{h}_{1, c}$, to generate $\bm{z}_1 = [\bm{z}_{1, 1}^\top, \bm{z}_{1, 2}^\top, \ldots, \bm{z}_{1, C}^\top]^\top$, where $\bm{z}_{1, c} \in \mathbb{R}^{2}$. The codeword $\bm{z}_1 \in \mathbb{R}^{2C}$ can be interpreted as the codeword of a rate-1/$2$ channel code. 

We then modulate the latent vectors, ${\bm{z}}_1$, to a complex-valued codeword, $\tilde{\bm{x}}_1 \in \mathbb{C}^C$. The modulation process is achieved by taking the first $C$ elements of ${\bm{z}}_1$ as the real part and the remaining elements as the imaginary part:
\begin{equation}
    \tilde{\bm{x}}_1 = {\bm{z}}_{1, 1:C} + j{\bm{z}}_{1, C:2C}.
    \label{equ:modulation}
\end{equation}
In this way, we can generate $B$ such codewords, $\tilde{\bm{x}}_d \triangleq [\tilde{\bm{x}}_1^\top, \ldots, \tilde{\bm{x}}_B^\top]^\top$, where $\tilde{\bm{x}}_d \in \mathbb{C}^{N_b}$ denotes the data symbols.

\subsubsection{Pilot symbol generation}
The pilot symbols in the proposed LISAC framework can either be fixed with unitary entries as in \eqref{eq:est_csi}, or obtained via end-to-end learning. Inspired by \cite{learn_a_pilot}, we consider learned pilot generation, which is shown to significantly improve the overall system performance.

Similarly to the generation process of the data symbols, $\tilde{\bm{x}}_d$, the LConv encoder shown in Fig. \ref{fig:jsc_framework} is adopted to produce the pilot symbols, $\tilde{\bm{x}}_p \in \mathbb{C}^{KM_p}$. We feed the fixed length-$KM_p$ vector $\bm{1}_{KM_p}$ into the LConv encoder of the pilot encoding function, $f_p(\cdot)$:
\begin{equation}
    \tilde{\bm{x}}_p = f_p(\bm{1}_{KM_p}).
\end{equation}

Next, we apply power normalization, expressed as:
\begin{align}
\label{equ:lisac_pn}
{\bm{x}} &= \frac{\tilde{\bm{x}} - \mu}{\sigma}, \\
 \mu \triangleq \frac{1}{KM}\mathbb{E}\left({\sum_i\tilde{{x}}[i]}\right)&; \; \sigma^2 \triangleq \frac{1}{KM}\mathbb{E}\left({\sum_i |\tilde{{x}}[i] - \mu|^2}\right), \notag
\end{align}
where $\tilde{\bm{x}} \in \mathbb{C}^{KM}$ is obtained by concatenating the pilot and data symbols, $\tilde{\bm{x}} \triangleq [\tilde{\bm{x}}_p^\top, \tilde{\bm{x}}_d^\top]^\top$. $\mu, \sigma$ are calculated by averaging over sufficient number of generated codewords during training.

\subsubsection{OFDM symbol mapping} \label{sec:ofdm_sym_map}
\begin{figure}[t]
     \centering
         \includegraphics[width=\columnwidth]{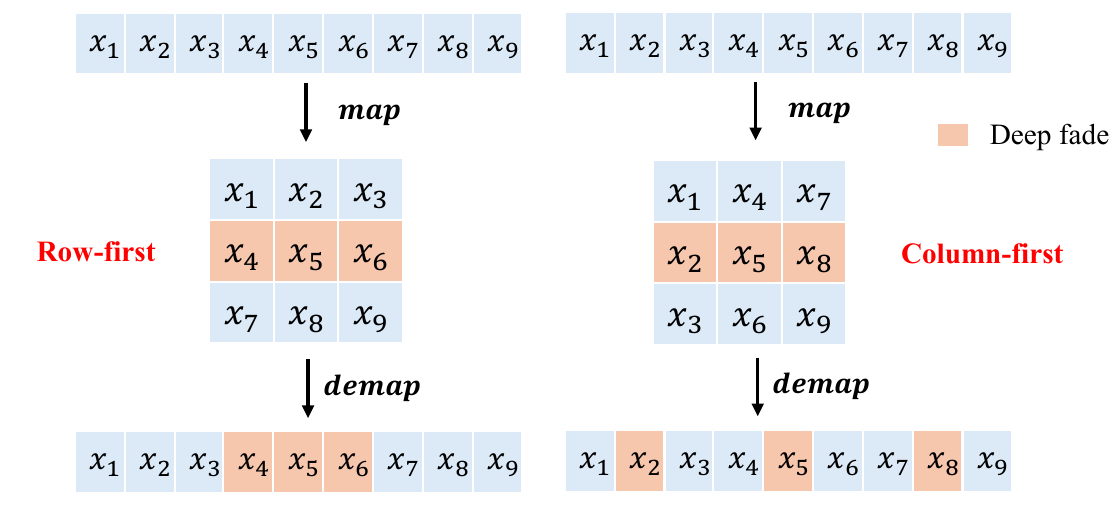}  
  \caption{Illustrations of the row-first and column-first OFDM symbol mapping strategies. It can be seen that when there exists a deep fade at a certain frequency band, the column-first strategy avoids consecutive data symbols with poor received SNR values after demapping.}
  \label{fig:mapping}
\end{figure}

Finally, we map the power normalized signal, $\bm{x}$, to $M$ OFDM symbols to form the transmitted signal, $\bm{X} \in \mathbb{C}^{K\times M}$, which can be expressed as:
\begin{equation}
    \bm{X} \leftarrow \textit{OFDMSymMap}(\bm{x}).
    \label{eq:ofdm_sym_map}
\end{equation}
Similarly to \cite{lisac_conf}, we consider two different OFDM symbol mapping strategies, where the first one loads the elements of $\bm{x}$  in a row-first fashion while the second one performs column-first loading. Note that the data mapping strategies are separately applied to the pilot and data OFDM symbols. The two strategies are illustrated in Fig. \ref{fig:mapping} for $K = M_d = 3$, and the corresponding power normalized data OFDM signal is of length-9. It is found in \cite{lisac_conf} that the two mapping strategies yield nearly identical performance over an AWGN channel, but the column first mapping strategy is strictly superior in the multi-path fading scenario. 

This phenomenon is explained in Fig. \ref{fig:mapping}: due to the frequency selective nature of the multi-path fading channel, some OFDM subcarriers  might experience deep fading (e.g., the second OFDM subcarrier in the figure). When the row-first data loading strategy is adopted, $K$ consecutive data symbols of poor quality are fed into the LConv decoder. The column-first strategy, on the other hand, distributes the $K$ symbols with poor received SNR across the block, which improves the overall decoding performance. This aligns with the principle of bit interleaved coded modulation \cite{bicm}, which is shown to be robust against burst channel errors. Numerical comparison of the two strategies are performed in the simulation section which corroborates the above arguments.

\subsection{LISAC receiver} 
Next, we present the details of the LISAC receiver illustrated on the RHS of Fig. \ref{fig:system_model}. 
The overall received signal, $\bm{Y}_c = [\bm{Y}_p, \bm{Y}_d]$ is comprised of the received pilot and data symbols. We first generate the MMSE estimate of the CSI, $\hat{\bm{h}}$, using $\bm{Y}_p$ and the pilot $\bm{X}_p$:
\begin{equation}
    \hat{\bm{h}} = (\widetilde{\bm{X}}_p^\dagger \widetilde{\bm{X}}_p + \sigma_c^2 \bm{I}_K)^{-1} \widetilde{\bm{X}}_p\bm{y}_p,
    \label{eq:est_csi_plt}
\end{equation}
where $$\widetilde{\bm{X}}_p \triangleq [\text{diag}(\bm{X}_p[:,1]), \ldots, \text{diag}(\bm{X}_p[:,M_p])]^\top,$$ and ${\bm{y}}_p \triangleq \text{vec}(\bm{Y}_p)$. We  note that $\hat{\bm{h}} = \hat{\bm{h}}(\bm{X}_p)$, i.e., the channel estimation output is a function of the pilot generated by the pilot encoding function, $f_p(\cdot)$.
After obtaining $\hat{\bm{h}}$, we perform channel equalization to generate the estimate of the transmitted data symbols, $\bm{X}_d$. The channel equalization process is comprised of the standard MMSE equalization to generate a coarse estimate, $\widetilde{\bm{X}}_d$ according to \eqref{eq:equ_xki}, followed by a neural network module, called ResidualNet, which further calibrates  $\widetilde{\bm{X}}_d$. 

In particular, we concatenate the estimated CSI, $\hat{\bm{h}}$, with each column of the received signal, $\bm{Y}_d$,  to form a complex-valued tensor with dimensions ${2\times K \times M_d}$, which is then converted to a real-valued tensor, $\widetilde{\bm{Z}} \in \mathbb{R}^{4\times K \times M_d}$ as follows:
$$\widetilde{\bm{Z}}[:, k, m] = [\Re({Y}_d[k, m]), \Im({Y}_d[k, m]), \Re(\hat{{h}}_k), \Im(\hat{{h}}_k)]^\top,$$
where $k\in [1, K], m \in [1, M_d]$. We then demap $\widetilde{\bm{Z}}$ to $B$ blocks, where the $b$-th block, $\widetilde{\bm{Z}}_b \in \mathbb{R}^{4\times C}$ is fed to the ResidualNet to produce the output, $\widetilde{\bm{X}}^r_b \in \mathbb{R}^{2\times C}$:
\begin{align}
    \{\widetilde{\bm{Z}}_b\}_{b\in [1, B]} &= \textit{OFDMSymDeMap}(\widetilde{\bm{Z}}), \notag \\
    \widetilde{\bm{X}}^r_b &= \text{Res}(\widetilde{\bm{Z}}_b),
    \label{eq:demap_res}
\end{align}
where $\text{Res}(\cdot)$ represents the processing of the ResidualNet which follows the same neural network architecture of the LConv encoder.
Note that the demapping process mirrors the OFDM symbol mapping in the LISAC transmitter.  By converting each of the  output blocks to a complex-valued signal and combining them together, we obtain the residual signal, ${\bm{X}}^r \in \mathbb{C}^{K\times M_d}$. The estimate of the transmitted signal, $\hat{\bm{X}}_d$ is obtained as 
\begin{equation}
    \hat{\bm{X}}_d = \widetilde{\bm{X}}_d + {\bm{X}}^r,
\end{equation}
which is fed to the LISAC decoder for further processing.

Similar with \eqref{eq:demap_res}, the equalized signal $\hat{\bm{X}}_d$ is also demapped to $B$ blocks, $\{\hat{\bm{x}}_{d, b}\}_{b\in [1, B]} = \textit{OFDMSymDeMap}(\hat{\bm{X}}_d)$. Each of the symbols, $\hat{\bm{x}}_{d, b}$, is demodulated to a real-valued matrix, $\hat{\bm{X}}_{d, b} \in \mathbb{R}^{2\times C}$, by concatenating the real and imaginary parts of $\hat{\bm{x}}_{d, b}$. The real-valued matrix is further fed to the LConv decoder, denoted by $g_d(\cdot)$, to generate the probability vector, $\bm{p}_b \triangleq [{p}_{b, 1}, \ldots, {p}_{b, C}] \in (0, 1)^C$:
\begin{equation}
    \bm{p}_b \triangleq g_d(\hat{\bm{X}}_{d, b}).
    \label{eq:lisac_dec}
\end{equation}
LConv decoder has a similar neural network architecture to LConv encoder, and is comprised of bi-GRU blocks, FC layers, and Sigmoid activation functions. However,  its bi-GRU blocks take 2d vectors as input and its FC layer produces scalar outputs. We denote the overall neural network parameters at the receiver as $\Theta_r$ which is comprised of the parameters of the ResidualNet and the LConv decoder.

\begin{remark}
    For a better understanding concerning the optimization process of the pilot encoding function, $f_p(\cdot)$, parameterized by $\bm{\Theta}_{t,p}$, we explicitly highlighted the backpropagation of the gradients, denoted by $\bm{g}(\bm{\Theta}_{t,p})$ w.r.t. $\bm{\Theta}_{t,p}$, using dashed lines in Fig. \ref{fig:system_model}. The gradient, comprised of the communication and the sensing part is expressed as:
    \begin{equation}
       \bm{g}(\bm{\Theta}_{t,p}) \leftarrow \underbrace{\frac{\partial \mathcal{L}_c}{\partial \hat{\bm{X}}_d} \frac{\partial \hat{\bm{X}}_d}{\partial \hat{\bm{h}}} \frac{\partial \hat{\bm{h}}(\bm{X}_p)}{\partial {\bm{X}}_p} \frac{\partial {\bm{X}}_p}{\partial {\bm{\Theta}}_{t,p}}}_{\text{comm.}} +  \underbrace{\frac{\partial \mathcal{L}_s}{\partial {\bm{Y}}_s} \frac{\partial {\bm{Y}}_s}{\partial {\bm{X}}_p}\frac{\partial {\bm{X}}_p}{\partial {\bm{\Theta}}_{t,p}}}_{\text{sense}},
    \end{equation}
    where the estimated CSI, $\hat{\bm{h}} = \hat{\bm{h}}(\bm{X}_p)$, is a function of the pilot signal as illustrated in \eqref{eq:est_csi_plt}.
\end{remark}

\begin{algorithm}
	\caption{Training algorithm for the LISAC transmitter and receiver with $E_o$ as the sensing loss.}
    \SetKwInOut{Input}{Input}
    \SetKwInOut{Output}{Output}
    \SetKwFunction{LConvEncode}{LConvEncode}
    \SetKwFunction{LConvDecode}{LConvDecode}
    \SetKwFunction{Modulate}{Modulate}
    \SetKwFunction{DeModulate}{DeModulate}
    \SetKwFunction{OFDMSymMap}{OFDMSymMap}
    \SetKwFunction{OFDMSymDeMap}{OFDMSymDeMap}
    \label{algorithm:train_lisac}
	
	\Input{$K, M_p. M_d, B, C, \lambda, N_e, N_d, N_{epoch}$}
        \Output{Optimized $\{\bm{\Theta}_t, \bm{\Theta}_r\}$}
	
	\BlankLine

	\For{$n=1$ \KwTo {$N_{epoch}$}}{
        Fix LISAC receiver parameters, $\bm{\Theta}_r$, \\
        \For{$n_t=1$ \KwTo $N_t$}{
            
            Randomly sample $ \bm{b}_b \in \{0, 1\}^{C}, \forall b \in [1, B]; v_\epsilon \sim \mathcal{U}(-1/2, 1/2)$,\\
            Set $(n_0, v_0) = (0, 0)$,\\
             $\tilde{\bm{x}}_d = f_s(\{\bm{b}_b\}_{b\in [1,B]}),$ \\
             $\tilde{\bm{x}}_p = f_p(\bm{1}_{KM_p}),$ \\
             $\bm{x} \leftarrow  \textit{Eqn.} \; \eqref{equ:lisac_pn},$  \Comment{{Power normalize}.} \\
             $\bm{X} \leftarrow \OFDMSymMap(\bm{x}),$ \\ \Comment{\textbf{LISAC transmitter processing}.} \\

            $(\bm{Y}_c, \bm{Y}_s) \leftarrow \textit{Eqn.} \; 
 \eqref{equ:comm_channel}, \; \textit{Eqn.} \; \eqref{equ:sense_y}$ \\
            \Comment{\textbf{Communication \& sensing channel.}} \\

            $\hat{\bm{h}} = (\widetilde{\bm{X}}_p^\dagger \widetilde{\bm{X}}_p + \sigma_c^2 \bm{I}_k)^{-1} \widetilde{\bm{X}}_p\bm{y}_p$; \\ \Comment{Channel estimation, \textit{Eqn.} \eqref{eq:est_csi_plt}.} \\
            $\widetilde{{X}}_d[k, m] =  \hat{{h}}_k {Y}_d[k, m] / (|\hat{{h}}_k|^2 + \sigma_c^2),$ \\
            $\hat{\bm{X}}_d = \widetilde{\bm{X}}_d + {\bm{X}}^r,$ where ${\bm{X}}^r \leftarrow$ \textit{Eqn.} $\eqref{eq:demap_res}$ \\ 
            \Comment{{Residual-assisted equalization}.} \\
             $\{\hat{\bm{x}}_{d, b}\}_{b\in [1, B]} \leftarrow \OFDMSymDeMap(\hat{\bm{X}}_d),$ \\
             $\hat{\bm{X}}_{d, b} \leftarrow \DeModulate(\hat{\bm{x}}_{d, b}),$\\
             ${\bm{p}}_b \leftarrow g_d(\hat{\bm{X}}_{d, b}),$\\
             \Comment{\textbf{LISAC receiver processing}.}\\

             $\mathcal{L}_c   \leftarrow \textit{Eqn.} \; \eqref{equ:bceloss};$ \\ $\mathcal{L}_s \leftarrow \textit{Eqn.} \; \eqref{equ:mse_sense_loss}$ using $\bm{Y}_s, v_\epsilon$. \\
             Optimize $\bm{\Theta}_t$ 
 using $\mathcal{L} = \mathcal{L}_c + \lambda \mathcal{L}_s$.
		}
            Fix LISAC transmitter parameters, $\bm{\Theta}_t$, \\ 
        \For{$n_r=1$ \KwTo $N_r$}{
            Generate $\{\bm{b}_b, \bm{p}_b\}_{b\in [1, B]}$ repeating line 4-20. \\
            $\mathcal{L}_c \leftarrow \textit{Eqn.} \; \eqref{equ:bceloss}.$\\
             Optimize $\bm{\Theta}_r$ 
 using $\mathcal{L}_c$.  \\
		}    
	}

	\BlankLine
\end{algorithm}

\subsection{Loss functions}
Next, we introduce the loss functions for both communication and sensing.

\subsubsection{Communication loss}
To optimize the encoder and decoder parameters to achieve a low BER value, we adopt the widely used binary cross entropy (BCE) loss \cite{kim2018communication}, which can be expressed as:
\begin{equation}
    \mathcal{L}_c = - \sum_{b}\sum_{c} \left({b}_{b,c} \log_2({p}_{b,c}) + (1-{b}_{b,c})\log_2(1-{p}_{b,c})\right),
    \label{equ:bceloss}
\end{equation}
where ${p}_{b, c}\in (0, 1)$ is the LISAC decoder output corresponding to the $c$-th bit of the $b$-th block. 

\subsubsection{Sensing loss adopting outlier MSE}
{
The loss function for target sensing is more complicated.
A naive solution is to directly adopt the outlier MSE, $E_o$, defined in \eqref{equ:outlier_mse}. However, the acquisition of the indices, $(\hat{n}_0, \hat{v}_0)$, involves argmax operation in \eqref{equ:ml_detect}, which is non-differentiable. To this end, we consider a proxy loss function, $\mathcal{L}_s$,  expressed as:
\begin{equation}
    \mathcal{L}_s = \mathbb{E}_{v_\epsilon} \sum_{(n, v) \neq (0, 0)} (n^2 + v^2)P_{(0,0) \rightarrow (n,v)}(v_\epsilon),
    \label{equ:mse_sense_loss}
\end{equation}
where we assume the ground truth $(n_0, v_0) = (0, 0)$ and $P_{(0,0) \rightarrow (n, v)}(v_\epsilon)$ is the pairwise outlier probability, defined following the maximum likelihood principle:
\begin{subequations} 
\begin{align}
    P_{(0,0) \rightarrow (n, v)}&(v_\epsilon) \triangleq P(|\gamma_{n,v,v_\epsilon}| > |\gamma_{0,0,v_\epsilon}|), \label{eq:pairwise_prob_a} \\
    &\approx \exp\left( - \frac{|r_{0,0, v_\epsilon}|}{2 \sigma^2} \right) I_0\left(\frac{|r_{n,v, v_\epsilon}|}{2 \sigma^2}\right)/2, \label{eq:pairwise_prob_b}\\
    &\le \exp\left( - \frac{|r_{0,0, v_\epsilon}| - |r_{n,v, v_\epsilon}|}{2 \sigma^2} \right)/2,
    \label{eq:pairwise_prob_c}
\end{align}
\end{subequations}
where $I_0(\cdot)$ denotes the Bessel function, $\sigma^2 = 
\frac{1}{\mathrm{SNR}_s}  = \sigma^2_s/A^2$ according to \eqref{eq:sense_snr}, while the cross correlation and discrete ambiguity terms, $\gamma_{n,v,v_\epsilon}$ and $r_{n,v,v_\epsilon}$ are both defined in \eqref{equ:ml_detect}. {The detailed derivations of \eqref{eq:pairwise_prob_a}, \eqref{eq:pairwise_prob_c} are given in \cite{isac_waveform_design}, and  omitted here due to page limitation.} 
Compared with the outlier MSE in \eqref{equ:outlier_mse}, $\mathcal{L}_s$ provides a `soft' measurement on how the estimated delay and Doppler indices, $(\hat{n}, \hat{v})$, deviate from the ground truth. Note that, even though $\mathcal{L}_s$ is not exactly the outlier MSE, we can still achieve satisfactory performance as we will show in Section \ref{sec:experiment}.

\subsubsection{CRLB as an alternative sensing loss}
We also consider training the LISAC model adopting CRLB as the sensing loss. In particular, the CRLB values of the delay, $\tau$, and Doppler, $f_D$, are calculated and their sum serves as the sensing loss.
Similar with the outlier MSE, the CRLB is also determined by the SNR of the sensing link, $\mathrm{SNR}_s$ and the power distribution over different OFDM symbols and subcarriers, $p_{k,m}$.
The CRLB values of $\tau$ and $f_D$ are derived in \cite{ofdm_crlb}. Since different OFDM systems would adopt different setups (e.g., bandwidth, $B$, number of subcarriers, $K$, and guard intervals, $K_G$), the authors in \cite{ofdm_crlb} take different delay and Doppler resolutions, $\Delta \tau$ and $\Delta f_D$, into account. 
In our case, we are interested in providing CRLB values that are independent of the resolutions $\Delta \tau$ and $\Delta f_D$, thus, we normalize the delay, $\tau$, and Doppler, $f_D$ using $\Delta \tau$ and $\Delta f_D$, and  the resulting CRLB of the normalized parameters, $(n^*, v^*)$, are given as:
\begin{align}
    \text{CRLB}(n^*) &= \frac{1}{8\pi^2 \mathrm{SNR}_s h_{\tau}(\bm{P})}, \notag \\
    \text{CRLB}(v^*) &= \frac{1}{8\pi^2 \mathrm{SNR}_s h_{f_D}(\bm{P})}, 
    \label{eq:crlb_express}
\end{align}
where $h_{\tau}(\bm{P})$ and $h_{f_D}(\bm{P})$ are expressed as:
\begin{align}
    h_{\tau}(\bm{P}) = \sum_{k=0}^{K-1} \sum_{m=0}^{M-1} \bar{k}^2p_{k,m} - \frac{\left(\sum_{k=0}^{K-1} \sum_{m=0}^{M-1} 
\bar{k} \bar{m} p_{k,m}\right)^2}{\sum_{k=0}^{K-1} \sum_{m=0}^{M-1}\bar{m}^2 p_{k,m}}, \notag \\
h_{f_D}(\bm{P}) = \sum_{k=0}^{K-1} \sum_{m=0}^{M-1} \bar{m}^2p_{k,m} - \frac{\left(\sum_{k=0}^{K-1} \sum_{m=0}^{M-1} 
\bar{k} \bar{m} p_{k,m}\right)^2}{\sum_{k=0}^{K-1} \sum_{m=0}^{M-1}\bar{k}^2 p_{k,m}},
\end{align}
where $\bar{k} = k - K/2, \bar{m} = m - M/2$ are the mean shifted indices w.r.t. the original indices, ${k\in [0,K-1], m\in [0,M-1]}$, and ${P}[k,m] = p_{k,m}$. We refer the detailed derivation of the CRLB expression to Appendix \ref{sec:APPA}. The loss function using CRLB, denoted as $\mathcal{L}_s^c$, is then expressed as:
\begin{equation}
    \mathcal{L}_s^c = \text{CRLB}(n^*) + \text{CRLB}(v^*).
    \label{eq:crb_loss}
\end{equation}

\subsection{Training of the LISAC system}
Combining the communication and sensing losses, the overall loss functions based on the outlier MSE and the CRLB metrics can be expressed as:
\begin{equation}
    \mathcal{L} = \left\{
    \begin{aligned}
    \mathcal{L}_c + \lambda \mathcal{L}_s  & , & E_o, \\
    \mathcal{L}_c + \lambda_c \mathcal{L}_s^c & , & \text{CRLB},
    \end{aligned}
    \right.
    \label{equ:total_loss}
\end{equation}
where $\lambda$ and $\lambda_c$ determine whether the learned waveform is sensing-oriented or communication-oriented. In particular, with a large $\lambda$/$\lambda_c$ value, one would expect a satisfactory outlier MSE/CRLB performance with a less satisfactory BER value, and vice versa for a small $\lambda$/$\lambda_c$ value.

To avoid getting stuck in local optima, we train the neural network parameters, $\bm{\Theta}_t$ and $\bm{\Theta}_r$, of the LISAC transmitter and receiver alternatively. In particular, we optimize the parameters of the LISAC transmitter, $\bm{\Theta}_t$, for $N_t$ iterations using the overall loss function, $\mathcal{L}$ defined in \eqref{equ:total_loss}, with $\bm{\Theta}_r$ fixed. Then, we fix $\bm{\Theta}_t$, and train LISAC receiver using only the communication loss, $\mathcal{L}_c$, for $N_r$ iterations. In general, we set $N_r > N_t$ as the LISAC decoder is harder to optimize. A large batch size is adopted, which is essential for achieving a satisfactory decoding performance. The overall training algorithm with outlier MSE as the sensing loss is summarized in Algorithm \ref{algorithm:train_lisac}, which can be easily extended to the scenario with CRLB  loss.


\newcolumntype{Z}{>{\raggedright\arraybackslash}b{1.5cm}}
\newcolumntype{Y}{>{\centering\arraybackslash}X} 

\begin{table}[t]

\caption{{List of key variables.}}
\centering
 \begin{tabularx}{\linewidth}{| Z | Y |} 
 \hline

 Variable & Description \\  \hline

 $K, K_G$ & The number of OFDM subcarriers and guard subcarriers in one OFDM symbol.  \\  \hline

 $M_p, M_d, M$ & The number of pilot and data OFDM symbols, satisfying $M = M_p + M_d$.  \\  \hline
 {$L$} & Number of paths in the communication link.  \\  \hline
 {$B, C$} & The number of code blocks within one OFDM frame and the number of bits within the block, satisfying $BC = KM$.  \\  \hline

 $\tau,  f_D$ & The delay and Doppler parameters. \\  \hline 
 $\Delta\tau, \Delta f_D$ & The delay and Doppler resolutions achieved by the OFDM frame. \\  \hline 
 \makecell[l]{$(n^*, v^*)$; \\ $(n_0, v_0)$} & The normalized delay and Doppler parameters according to \eqref{equ:delay_doppler} and their integer parts. \\  \hline 
 $p_{k,m}$ & The power allocated to the $k$-th OFDM subcarrier of the $m$-th OFDM symbol. \\  \hline 
 \makecell[l]{$\gamma_{n,v, v_\epsilon}$, \\ $r_{n,v, v_\epsilon}$} & {The cross correlation and discrete ambiguity function for ML parameter estimation.} \\  \hline 
 $(\hat{n}_0, \hat{v}_0)$ & The ML estimates of $(n_0, v_0)$. \\  \hline 
 $E_o$ & The expected MSE between $(n_0, v_0)$ and $(\hat{n}_0, \hat{v}_0)$. \\  \hline 
 $\bm{X}, \bm{X}_p, \bm{X}_d$ & The OFDM frame, $\bm{X}$, is obtained by concatenating the pilot and data OFDM symbols, $\bm{X}_p$ and $\bm{X}_d$.\\  \hline 
 $\bm{h}, \hat{\bm{h}}(\bm{X}_p)$ & The ground truth CSI and its MMSE estimate using pilot signal, $\bm{X}_p$. \\  \hline
 $\hat{\bm{X}}_d, \widetilde{\bm{X}}_d, {\bm{X}}^r$ & The equalized signal, $\hat{\bm{X}}_d$, is obtained by summing up the MMSE equalization output, $\widetilde{\bm{X}}_d$, and the ResidualNet output, ${\bm{X}}^r$. \\  \hline
 $\bm{\Theta}_t,  \bm{\Theta}_r$ & The neural network parameters of the LISAC transmitter and receiver.\\  \hline 
 $\mathcal{I}(\bm{\theta})$ & The Fisher information matrix of the delay and Doppler parameters to calculate the CRLB values. \\  \hline
 $\mathcal{L}_s, \mathcal{L}_s^c$ & The sensing losses corresponding to the outlier MSE, $E_o$ and the CRLB, respectively. \\  \hline
 $\lambda, \lambda^c$ & The hyper parameters for $\mathcal{L}_s$ and $\mathcal{L}_s^c$, respectively. \\  \hline
 \end{tabularx}
\label{tab:list_variables}

\end{table}

\section{Numerical Experiments}

\label{sec:experiment}
\subsection{Parameter Settings and Training Details}
Unless otherwise stated, we consider a rate-1/2 LISAC codeword and adopt the modulation scheme detailed in \eqref{equ:modulation} leading to a spectrum efficiency equals to one. 
Both the AWGN and multi-path fading channels are considered. For different numbers of OFDM subcarriers, $K$, and OFDM symbols, $M$, we vary the code length, $C$, and the number of bit blocks, ${B}$, to satisfy the condition, $BC = KM$. For the multi-path fading channel, we set the number of paths $L = 3$ and the number of pilot symbols $M_p = 2$. 

We set the number of RNN layers to 3 and the hidden dimension to $N_h = 256$ for the LISAC encoder and the ResidualNet, while they are set to $(5, 256)$ for the LISAC decoder. The batch size is $256$, and the learning rate for all the parameters is set to $10^{-3}$. Adam optimizer is adopted and the model is trained for $500$ epochs. For each epoch, the LISAC transmitter is optimized using $N_t = 3\times 10^4$ iterations, while the LISAC receiver is with $N_r = 5\times 10^4$ iterations. The LISAC transceiver are trained at fixed $\mathrm{SNR}_c$ and $\mathrm{SNR}_s$ but evaluated at different $\mathrm{SNR}_c$ and $\mathrm{SNR}_s$ values. 
The distribution of the input bit sequence is denoted by $P_{\bm{b}}$, where each input bit follows an i.i.d. Bernoulli distribution, $\mathcal{B}(1/2)$.
The rate-1/2 convolutional code with constraint length of 7 with polynomial generator, $(171, 133)$, is adopted as the baseline. For a fair comparison with the proposed LISAC scheme, the rate-1/2 convolutional code is QPSK modulated to provide a spectral efficiency equals to one.

\begin{figure}[!t]
\centering
\includegraphics[width=0.9\columnwidth]{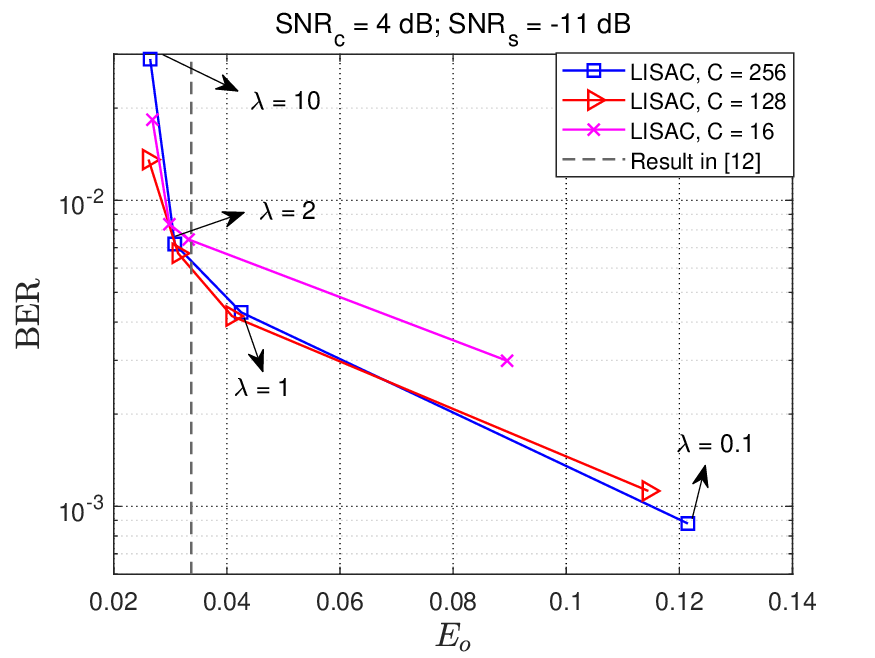}
\caption{Communication and sensing trade-off with different codeword lengths, $C$. Different trade-off points are achieved with varying $\lambda \in \{0.1, 1, 2, 10\}$. We also provide the performance achieved by  convolutional code using power allocation scheme described in \cite{isac_waveform_design}.}
\label{fig:fig_sctradeoff}
\end{figure}

\subsection{Performance Evaluation for AWGN Channel}
For an AWGN channel, there is no need to employ pilot symbols for channel estimation, i.e., $M_p = 0, M = M_d$. 

\begin{figure}[!t]
\centering
\includegraphics[width=0.95\columnwidth]{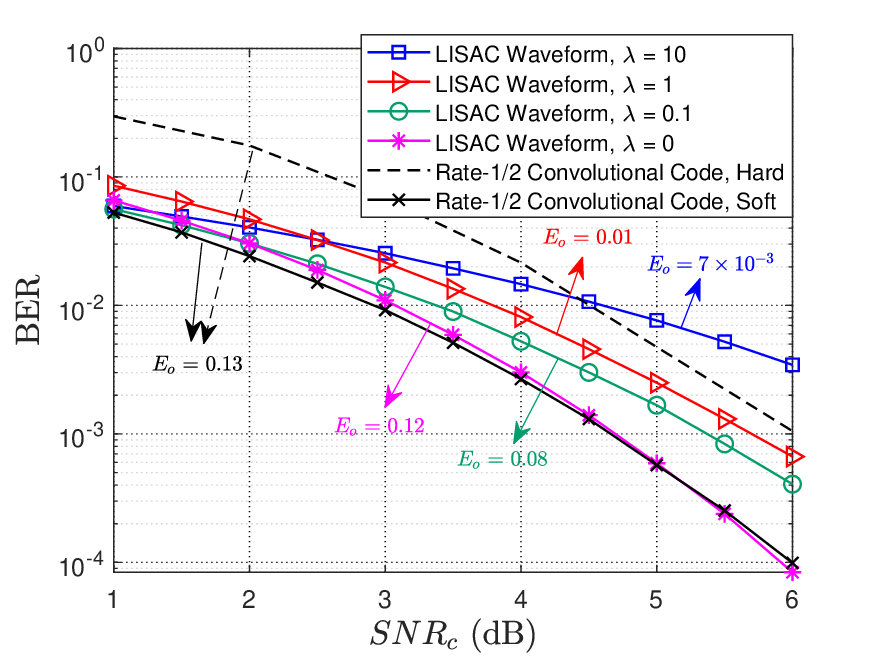}
\caption{The BER comparison between the LISAC waveforms trained with different $\lambda$ values and the 1/2-rate convolutional code using hard/soft QPSK demodulation.}
\label{fig:fig_ber_performance}
\end{figure}

\begin{figure}[!t]
\centering
\includegraphics[width=\columnwidth]{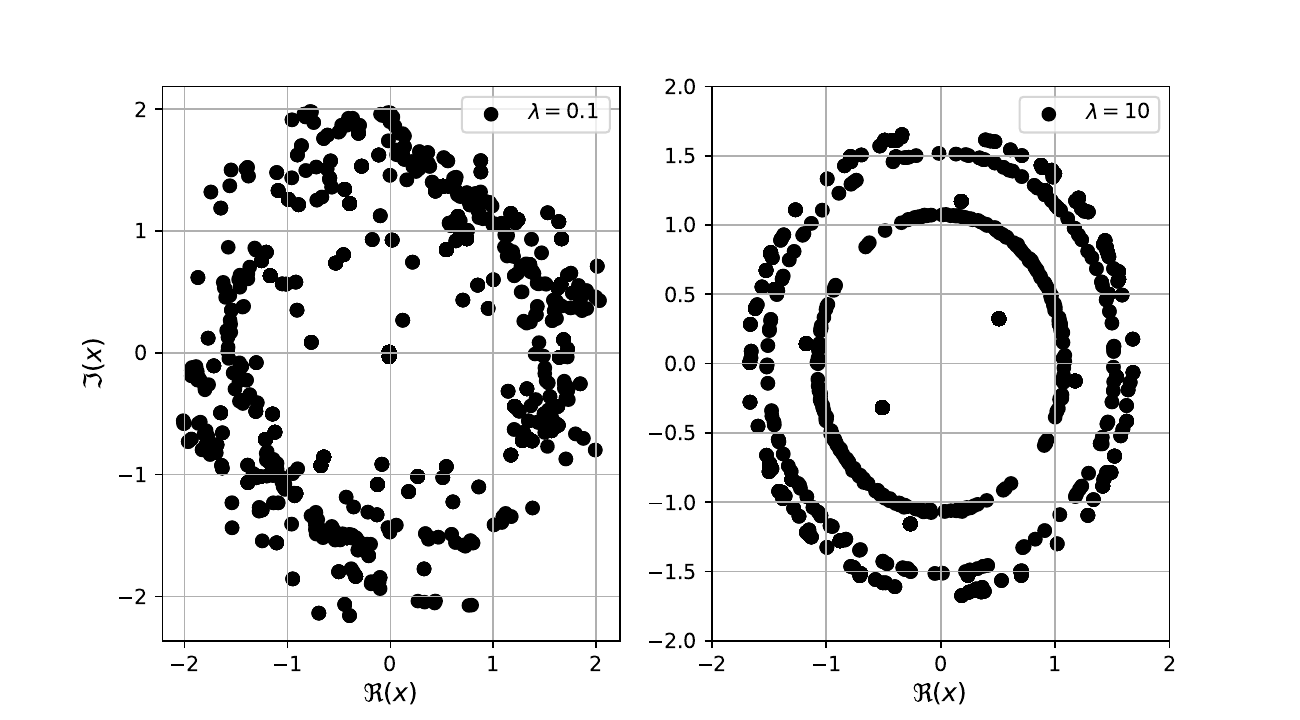}
\caption{Visualization of the learned codewords for block length, $C = 8$. The x/y-axes are the real/imaginary parts of $\bm{x}$, respectively. Visualizations for $\lambda = 0.1$ (left) and $\lambda = 10$ (right) are provided.}
\label{fig:fig_vis_waveform}
\end{figure}

\subsubsection{Communication-sensing trade-off} 
We first study the communication-sensing trade-off between BER and sensing MSE, $E_o$, defined in \eqref{equ:outlier_mse}. In this simulation, we consider the OFDM setting with $K = 32, M = 8$, while the communication and sensing SNRs are set to $\mathrm{SNR}_c = 4$ dB and $\mathrm{SNR}_s = -11$ dB, respectively. Different codeword lengths,  $C = \{16, 128, 256\}$ are considered which correspond to the number of blocks $B = \{16, 2, 1\}$. For a specific $(B, C)$ setup, we train four distinct LISAC models with different $\lambda \in \{0.1, 1, 2, 10\}$ values. As a result, $12$ models are obtained, each yields a distinct trade-off between BER and sensing MSE value as plotted in Fig. \ref{fig:fig_sctradeoff}.  The dashed line in the same figure represents the sensing performance achieved by the analytical optimization method presented in \cite{isac_waveform_design} where the authors adopt QPSK modulation and assign different powers to the OFDM symbols for satisfactory sensing performance.

We observe that the BER performance with a large $C$ is better by comparing the LISAC models with different $C$ parameters at $\lambda = 0.1$.  This is intuitive as the longer the block length the better the error correction ability. {Note that the waveform with $\lambda = 0.1$ puts more emphasis on communication performance.} However, when $\lambda$ becomes larger, e.g., $\lambda = 10$, the LISAC waveform with $C = 256$ yields the worst BER performance. This is attributed to the fact that, for $C = 256$, it is hard for the LISAC encoder to learn a length-$256$ codeword that is good for both sensing and error correction. The neural network is likely to get stuck at a local optima dominated by the sensing performance, instead of the global optima with a lower BER. When the block length reduces, e.g., $C = 128$, the LISAC encoder is more likely to reach the global optima.
We find in simulations that the BER performance of the baseline scheme in \cite{isac_waveform_design} is $0.13$, which is significantly outperformed by all the LISAC models due to the non-uniform power allocation over OFDM symbols \cite{isac_waveform_design}. It can be seen in the figure that the LISAC system with $\lambda > 2$ yields even smaller sensing MSE compared with the benchmark (specifically designed for radar sensing), which is plausible as the analytical optimization method may converge to a sub-optimal solution.

\subsubsection{BER comparison with a conventional coded modulation scheme}
Next, we consider the setting with $K = 32, M = 16$. The coding scheme adopts $C = 128, B = 4$. 
Four different models achieving different communication and sensing trade-offs are trained for $\lambda \in \{0, 0.1, 1, 10\}$ values and fixed sensing SNR, $\mathrm{SNR}_s = -13$ dB. The relative performance between the LISAC waveform and the conventional convolutional code with QPSK modulation evaluated for $\mathrm{SNR}_c \in [1, 6]$ dB is shown in Fig. \ref{fig:fig_ber_performance}. It is worth noting that two baseline schemes are considered, both employing the same coded modulation but differing in whether the QPSK demodulator outputs soft or hard log likelihood ratio (LLR) values.

As can be seen, the LISAC waveform with $\lambda = 0$ has the best BER performance yet the worst sensing performance, i.e., $E_o$, compared to the models trained with non-zero $\lambda$ values.  When $\lambda$ grows, the communication performance degrades yet the outlier MSE decreases. All the LISAC waveforms outperform the conventional convolution code baselines in terms of outlier MSE\footnote{We note that the LISAC model with $\lambda = 0$ outperforms the baseline scheme showing that uniform power constellation is a strictly sub-optimal solution.}. It is shown in the figure that the LISAC model with $\lambda = 0$ yields comparable BER performance with the baseline adopting soft LLR values yet achieving an approximately 0.5 dB SNR gain over the baseline with hard LLR values.

\subsubsection{Visualization of the learned codewords} 
Finally, we visualize the symbols employed by the learned codewords to provide more insights on the properties of the coded modulation scheme. In this simulation, we consider the same OFDM setup as in Fig. \ref{fig:fig_sctradeoff}. The block length is set to $C = 8$, corresponding to $2^C = 256$ different codewords\footnote{Though in practice, we adopt much larger $C$ for better error correction ability. In this simulation, we set $C = 8$ to simplify the visualization.}. Each of the $2^C$ codewords has $C = 8$ complex symbols and in total, $C2^C$ complex symbols are shown altogether in each of the plots in Fig. \ref{fig:fig_vis_waveform}. 
The two plots correspond to LISAC waveforms trained under $\lambda = \{0.1, 10\}$. Symbols corresponding to $\lambda = 0.1$ appear to be more spread across the I/Q plane, in order to reduce the BER by maximizing the distance between codewords. The codewords obtained with $\lambda = 10$, on the other hand, are concentrated on several co-centric circles. Since the elements belonging to the same circle have the same amplitude, information can only be embedded on the phase of the symbols. Thus, the waveform trained with $\lambda = 10$ has less satisfactory BER performance, as observed in Fig. \ref{fig:fig_sctradeoff}. On the other hand, there are only few possible amplitudes, $p_{k,m}$, available for the waveform with $\lambda = 10$ leading to a better sensing performance. This aligns with the intuition that deterministic amplitude signals are better suited for sensing.

\subsection{Performance over Multi-path Fading Channel}

\begin{figure}[!t]
\centering
\includegraphics[width=\columnwidth]{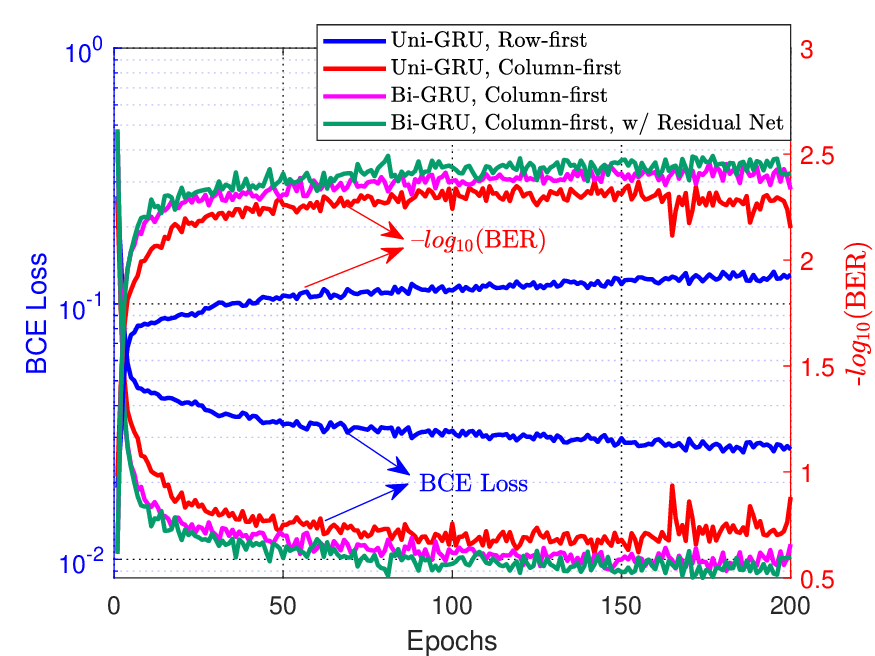}
\caption{The BCE loss and BER performance of the LISAC model in the validation phase over the first 200 epochs. We adopt an OFDM setup with $(K = 32, M = 20, M_p = 2)$, and train different LISAC models with different neural network architectures under a multi-path fading channel model with perfect CSI.}
\label{fig:bce_ber_epoch}
\end{figure}

We then consider a multi-path fading channel. Throughout this subsection, we set the number of OFDM subcarriers to $K = 32$, and the number of pilot and data OFDM symbols to $(M_p, M_d) = (2, 16)$, respectively. Thus, the number of bits in the input sequence, $\bm{b}$, equals to $576$ and we set $B = 1, C = 576$, i.e., the LISAC encoder takes the entire bit sequence, $\bm{b}$, as input.

We first show the effectiveness of the proposed neural network structures illustrated in Section \ref{sec:LISAC}. In this simulation, we only focus on the communication performance by setting $\lambda = 0$ and assume perfect CSI is available at the receiver, i.e., the channel estimation procedure in \eqref{eq:est_csi_plt} is skipped in this scenario. We show both the BCE loss and the BER performance (presented as $-\log(\text{BER})$) of the LISAC models in Fig. \ref{fig:bce_ber_epoch}, which are calculated in the validation phase over the first 200 epochs. As seen in the figure, the column-first OFDM symbol mapping strategy significantly outperforms the row-first counterpart for both BER and BCE performances. It is also observed that introducing bi-directional GRU to the LConv encoder and decoder help to improve the error correction ability of the learned waveform over the uni-directional one. We also find that adopting the ResidualNet further improves the decoding performance. Thus, we adopt column-first OFDM symbol mapping strategy and the ResidualNet for the remaining simulations.

\begin{figure*}
     \centering
     \begin{subfigure}{0.64\columnwidth}
         \centering
         \includegraphics[width=\columnwidth]{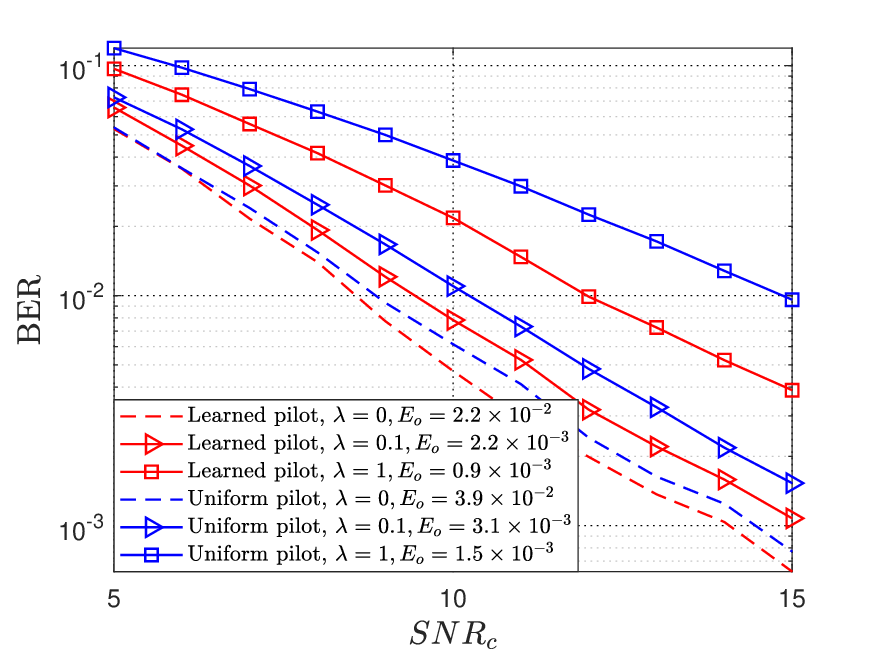}
         \caption{}
     \end{subfigure}
     \begin{subfigure}{0.64\columnwidth}
         \centering
         \includegraphics[width=\columnwidth]{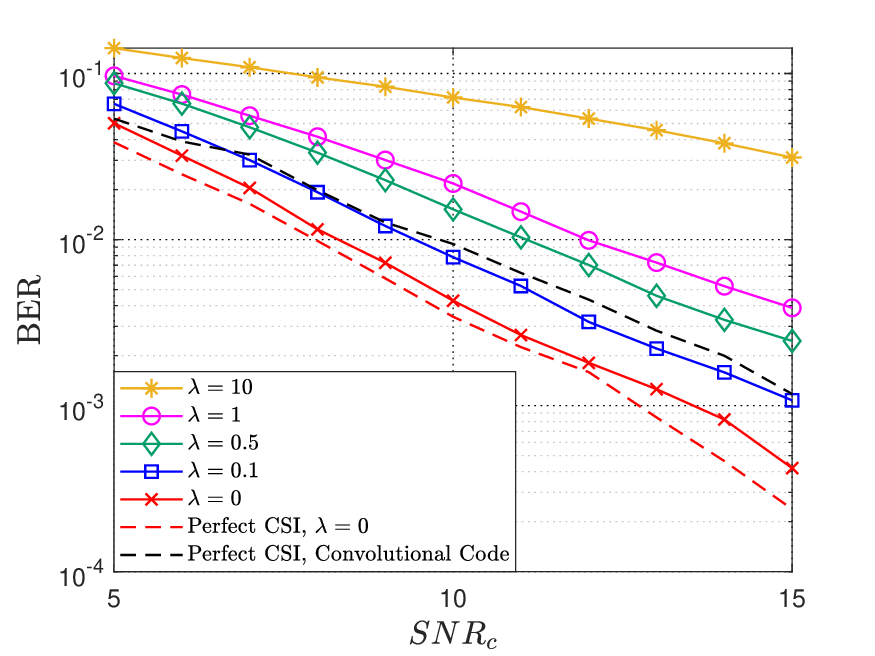}
         \caption{}
     \end{subfigure}
     \begin{subfigure}{0.64\columnwidth}
         \centering
         \includegraphics[width=\columnwidth]{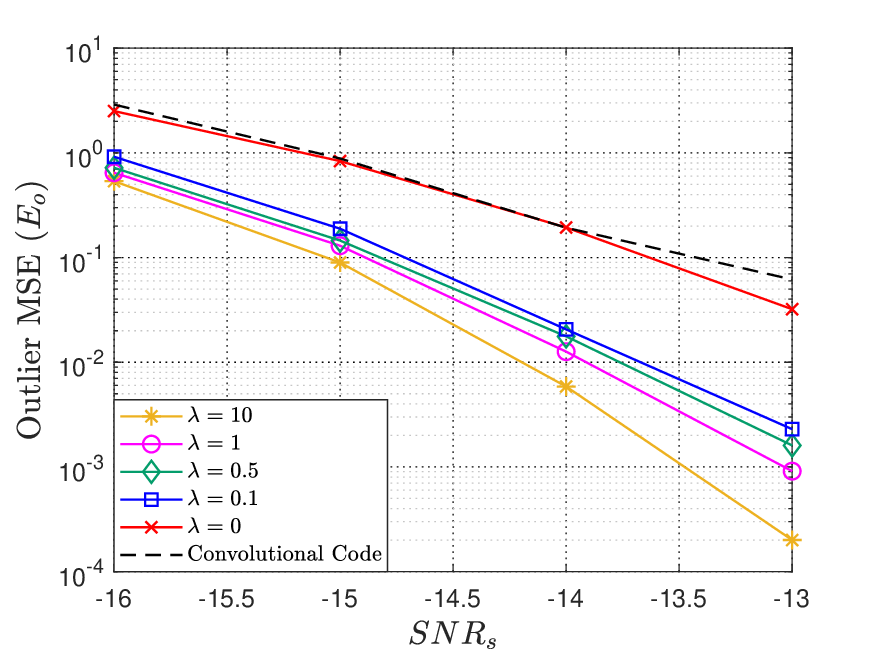}
         \caption{}
     \end{subfigure}
  \caption{Performance evaluation of the proposed LISAC models.  (a) Comparison between the LISAC models adopting pilots with unitary entries and those adopting learned pilot. (b) \& (c) Different LISAC models are trained with different $\lambda$ values to achieve different communication and sensing trade-offs.}
\label{fig:simu_576}
\end{figure*}

Next, we evaluate the sensing and communication performances of the proposed LISAC model. In this simulation, we consider varying SNR values for the communication link with $\mathrm{SNR}_c \in [5, 15]$ dB and fix the sensing SNR to $\mathrm{SNR}_s = -13$ dB.  To start with,  BER and sensing performances achieved by the LISAC model adopting pilots with unitary entries and  pilots produced by the pilot encoding function, $f_p(\cdot)$, are shown in Fig. \ref{fig:simu_576} (a). As can be seen, for all considered $\lambda$ values, LISAC model with learned pilots outperform those adopting uniform ones in terms of both $E_o$ and BER. When $\lambda = 0$, i.e., only the communication performance is taken into account, an approximately $0.5$ dB gain is observed when BER equals to $10^{-3}$. When $\lambda$ grows, i.e., the sensing performance is taken into account, the SNR gain of the model with learned pilots becomes even larger where the learned one outperforms the uniform one adopting $\lambda = 1$ by 3 dB when BER $= 10^{-2}$. We can conclude that adopting pilots with unitary entries in the first $M_p$ OFDM symbols is sub-optimal for sensing, thus, these LISAC models sacrifice their communication performances to achieve comparable sensing performances w.r.t. the learned ones. Based on the above analysis, we adopt LISAC models with learned pilots in the following simulations.

We then consider LISAC models trained with different $\lambda$ values together with the conventional baseline in Fig. \ref{fig:simu_576}(b) \& (c).
The conventional baseline adopts a rate-1/2 convolutional code with QPSK modulation. At the receiver, we first perform MMSE equalization assuming perfect CSI followed by QPSK demodulation and Viterbi decoding. Note that the adopted QPSK demodulator produces soft LLR outputs for the coded bit sequence to achieve a superior BER performance as in Fig. \ref{fig:fig_ber_performance} in the AWGN case.
It can be seen from Fig. \ref{fig:simu_576}(b) that the LISAC model learns to generate satisfactory channel estimation quality for subsequent channel equalization and decoding processes by comparing the curve with `$\lambda = 0$' with that using perfect CSI. We also observe an approximately $2.5$ dB SNR gain by comparing the LISAC model with $\lambda = 0$ and the baseline scheme adopting convolutional code both using perfect CSI when BER equals to $10^{-3}$. 

The sensing performance is evaluated in Fig. \ref{fig:simu_576}(c). It is shown that adopting constant amplitude waveform is strictly sub-optimal in this setup as the conventional scheme with QPSK modulation can only achieve comparable $E_o$ performance w.r.t. the LISAC model with $\lambda = 0$, a model which is not optimized for sensing performance at all. The LISAC models trained under non-zero $\lambda$ values can significantly outperform the conventional baseline.
Combining Fig. \ref{fig:simu_576}(b) \& (c), we can observe that the LISAC model trained with $\lambda = 0.1$ outperforms the baseline in terms of both BER and outlier MSE, showing the effectiveness of the proposed scheme.

\begin{figure*}
     \centering
     \begin{subfigure}{0.64\columnwidth}
         \centering
         \includegraphics[width=\columnwidth]{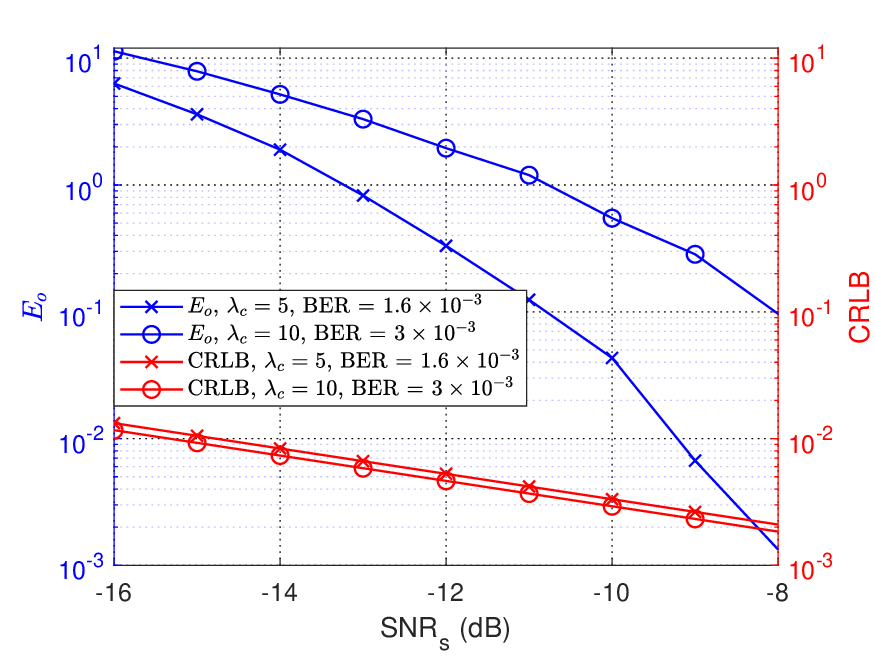}
         \caption{}
     \end{subfigure}
     \begin{subfigure}{1.28\columnwidth}
         \centering
         \includegraphics[width=\columnwidth]{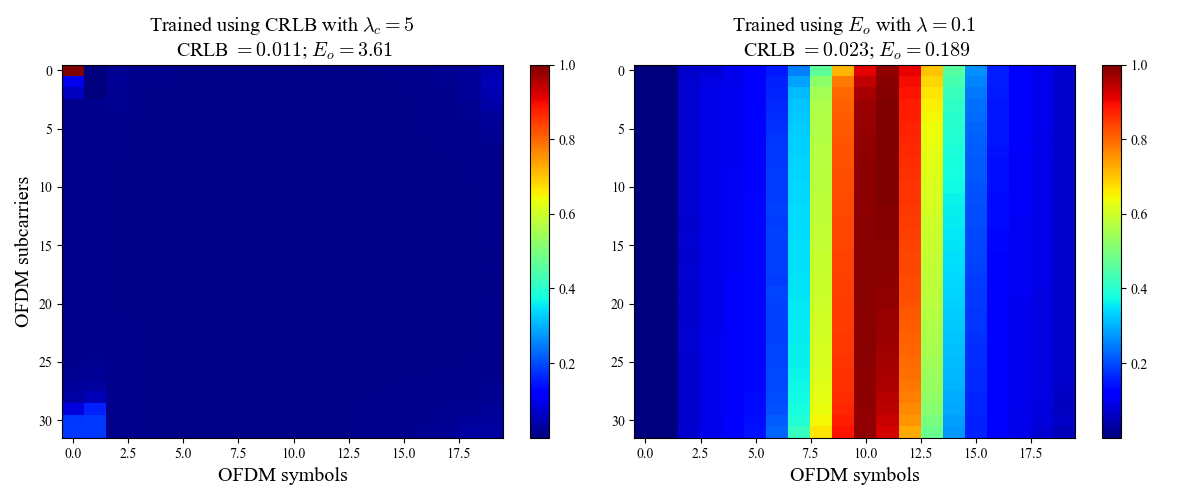}
         \caption{}
     \end{subfigure}

  \caption{Performance evaluation for the LISAC models trained with different sensing metrics, $\mathcal{L}_s$ and $\mathcal{L}_s^c$. (a) $E_o$ and CRLB performances obtained by LISAC models trained using $\mathcal{L}_s^c$ with different $\lambda_c \in \{5, 10\}$ values. (b) Power distributions ($p_{k,m}$) over different OFDM symbols/subcarriers obtained by two LISAC models evaluated under $(\mathrm{SNR}_c, \mathrm{SNR}_s) = (10, -15)$ dB: the left one is trained using CRLB with $\lambda_c = 5$ while the right one is trained using $E_o$ with $\lambda = 0.1$.}
\label{fig:simu_crlb}
\end{figure*}

{Finally, we consider the LISAC models  trained with the CRLB loss, i.e., $\mathcal{L}_s^c$ in \eqref{eq:crb_loss}, as the sensing loss function. The simulation setup is identical to that in Fig. \ref{fig:simu_576}. Two different LISAC models are trained with $\lambda_c \in \{5, 10\}$, aiming to achieve different trade-offs between BER and CRLB. Both models are evaluated under $\mathrm{SNR}_s \in [-16, -8]$ dB and $\mathrm{SNR}_c = 10$ dB.
As can be seen in Fig. \ref{fig:simu_crlb}(a), the LISAC model with a small $\lambda_c$ value achieves a better BER performance yet with a higher CRLB value. We also provide their outlier MSE ($E_o$) performances in the same figure. At a first glance, there is a large gap between the CRLB \texttt{-} the theoretical lower bound of the MSE \texttt{-} and the numerically simulated $E_o$, especially when $\mathrm{SNR}_s$ is low. This is because the ML estimator illustrated in Section \ref{sec:problem} makes hard decisions which breaks the unbiased estimator assumption of CRLB, failing to capture the MSE performance in low SNRs.
It is worth mentioning that when $\mathrm{SNR}_s = -8$ dB, the $E_o$ achieved by the LISAC model with $\lambda_c = 5$ is even smaller than the CRLB. This is due to the fact that the $E_o$ only evaluates the MSE of the integer parts, i.e., the ground truth, $(n_0, v_0)$ and the estimated $(\hat{n}_0, \hat{v}_0)$. The CRLB, on the other hand, also takes the fractional parts,  $(n_\epsilon, v_\epsilon)$, into account. We further note that the LISAC model with a superior CRLB performance corresponding to large $\lambda_c$ yields a high $E_o$ value. This is plausible as the two metrics are focusing on different aspects as analyzed above.

We further explore different transmit powers allocated to the OFDM symbols and subcarriers (i.e., $p_{k,m}$) for a more comprehensive understanding of the learned waveforms corresponding to different sensing loss functions. In practice, the power distribution is calculated by averaging over bit sequences:
\begin{equation}
    p_{k, m} = \mathbb{E}_{\bm{b\sim P_{\bm{b}}}}(|X[k, m]|^2),
\end{equation}
where $k\in [1, K], m\in [1, M]$, and $P_{\bm{b}}$ is the distribution of the input bit sequence. 
Power distributions of models trained with different metrics are visualized in Fig. \ref{fig:simu_crlb}(b).  In particular, the LISAC model trained using CRLB with $\lambda_c = 5$ and that trained using $E_o$ with $\lambda = 0.1$ are considered due to the fact that they achieve similar BER values equal to $0.008$ under $\mathrm{SNR}_c = 10$ dB. The sensing performances of both models are evaluated under $\mathrm{SNR}_s = -15$ dB. As seen in Fig. \ref{fig:simu_crlb}(b), for the model trained with CRLB, the power concentrates on the corners of the 2D grid, especially on the top left corner. For the model trained with $E_o$, on the other hand, most transmit power is allocated to the OFDM symbols located in the middle. The distinct preferred power allocation again indicates that using CRLB as loss function does not necessarily enhance outlier MSE as shown in Fig. \ref{fig:simu_crlb}(a).}

\section{Conclusion}
In this paper, we proposed LISAC, a deep learning-based coded waveform design for ISAC systems using OFDM. At the transmitter, the LISAC encoder employs the pilot and data encoding functions, both parameterized by RNNs, to enable a more flexible waveform design leading to improved sensing and communication performances. At the receiver, a residual-assisted MMSE equalizer is employed to enhance the estimation of the transmitted data symbols followed by an RNN-based decoder that efficiently decodes information bits. The proposed LISAC model is trained using a weighted loss function, and different sensing and communication trade-offs are achieved by adopting different weights. Simulation results demonstrate that the proposed LISAC model outperforms the baseline schemes under both AWGN and multi-path fading channels. 

Our study highlights two key challenges that merit further investigation:
\begin{itemize}
    \item The modulation scheme of the proposed LISAC model simply converts two real-valued symbols into a single complex-valued symbol achieving the same spectrum efficiency as the QPSK modulation. However, there exists scenarios where higher data rates/modulation orders are required. It is shown in \cite{turbo_mod} that learning-based modulators/demodulators,  parameterized by FC layers and trained along with the channel encoder/decoder, achieve satisfactory performance in the AWGN channel which can be a possible solution to the considered LISAC system.
    \item It is well known that the error correction ability of convolutional codes is less effective compared with  linear block codes, e.g., LDPC and Polar codes. This aligns with the observations in the paper, as LISAC exhibits high BER even under a large $\mathrm{SNR}_c$ value. To enhance the error correction ability of the LISAC waveform, a possible solution is to employ LDPC codes as an outer code forming a concatenated code \cite{ten2001convergence} along with the LISAC codeword. However, this requires the output of the LISAC decoder to be LLR values and is compatible with the  LDPC decoder which needs further investigation.
\end{itemize}

\bibliographystyle{IEEEbib}
\bibliography{refs}

\appendices
\section{On the Derivation of the CRLB in \eqref{eq:crb_loss}}\label{sec:APPA}
In this appendix, we provide detailed derivation of the CRLB in \eqref{eq:crb_loss}. 
To start with, we consider the normalized delay and Doppler parameters, $(n^*, v^*)$, illustrated in \eqref{equ:delay_doppler}, which are obtained by dividing the original delay and Doppler parameters, $\tau, \Delta f_D$, by their corresponding resolutions, $\Delta \tau, \Delta f_D$, respectively. 
Since the CRLB also takes the fractional delay component, $n_\epsilon$, into account, the time domain channel illustrated in \eqref{equ:td_y} is modified as:
\begin{equation}
    {y}[k, m] = {x}[k-n^*, m]e^{j2\pi\frac{v^*m}{M}} + {w}[k, m],
\end{equation}
where we omit the subscript of the sensing signal, $\bm{y}$, and set the complex gain, $a = 1$. 
Note that we set the indices\footnote{{It is also possible to adopt the indices, $k \in [1, K], m\in [1, M]$, leading to a different CRLB value. Instead of employing the indices starting from one, we adopt the mean shifted version following the literature \cite{ofdm_crlb, crlb_derive} to facilitate the analysis.}} to $k\in [-K/2, K/2-1], m\in [-M/2, M/2-1]$ and the sensing SNR to $\mathrm{SNR}_s = \frac{1}{\sigma^2_s}$. The frequency domain response can be expressed as:
\begin{align}
{Y}[k, m] = {X}[k, m]e^{j2\pi\frac{v^* m}{M}}e^{-j2\pi\frac{n^* k}{K}} + {W}[k, m], 
\label{equ:sense_y_crlb}
\end{align}
which can be written in a matrix form:
\begin{equation}
    \bm{Y}  = \bm{D}\bm{X}\bm{B} + \bm{W},
\end{equation}
where {$\bm{D} = \text{diag}([e^{-j2\pi\frac{n^* K/2}{K}}, \ldots,e^{-j2\pi\frac{n^*(K/2-1)}{K}}])$, $\bm{B} = \text{diag}([e^{j2\pi\frac{v^*M/2}{M}}, \ldots,e^{j2\pi\frac{v^*(M/2-1)}{M}}])$.} 
The likelihood function of the parameters, $\bm{\theta} \triangleq [n^*, v^*]$, is presented as:
\begin{equation}
    {P(\bm{y}; \bm{\theta}) = \frac{1}{\sigma_s^2 \pi^{KM}} \exp(-\frac{(\bm{y} - \bm{s})^\dagger (\bm{y} - \bm{s})}{\sigma^2_s}),}
    \label{eq:likelihood}
\end{equation}
where $\bm{y}, \bm{s} = \text{vec}(\bm{Y}), \text{vec}(\bm{D}\bm{X}\bm{B})$.

With \eqref{eq:likelihood}, the Fisher information matrix, $\mathcal{I}(\bm{\theta})_{i,j}, i, j \in [1, 2]$, can be expressed as:
\begin{equation}
    \mathcal{I}(\bm{\theta})_{i,j} = -\mathbb{E}\left( \frac{\partial^2  \log P(\bm{y}; \bm{\theta})}{\partial \theta_i \partial \theta_j} \right) = \frac{2}{\sigma_s^2} \Re\left( \frac{\partial \bm{s}^\dagger}{\partial \theta_i} \frac{\partial \bm{s}}{\partial \theta_j} \right),
\end{equation}
where we use the fact that $\mathbb{E}\left(\bm{y}-\bm{s}\right) = \bm{0}$ for the second equation.
The exact formula for the partial derivatives, $\frac{\partial \bm{s}}{\partial \theta_i}$, can be expressed as:
\begin{align}
    \frac{\partial \bm{s}}{\partial \theta_1} &= -j2\pi \frac{1}{K}[\underbrace{-K/2, \ldots, K/2-1}_{\text{repeat $M$ times}}]^\top \odot \bm{s}, \notag \\
    \frac{\partial \bm{s}}{\partial \theta_2} &= -j2\pi \frac{1}{M}[\underbrace{-M/2}_{\text{repeat $K$ times}}, \ldots, \underbrace{M/2-1}_{\text{repeat $K$ times}}]^\top \odot \bm{s},
    \label{eq:partial_derive}
\end{align}
where $\theta_1, \theta_2$ denote the delay and Doppler parameters, $n^*, v^*$, respectively.
With \eqref{eq:partial_derive}, the CRLB for each element, $\mathcal{I}(\bm{\theta})_{i,j}, i, j \in [1, 2]$ is:
\begin{align}
    \mathcal{I}(\bm{\theta})_{1, 1} = \frac{8\pi^2 \mathrm{SNR}_s}{K^2} \sum_{k=0}^{K-1} \sum_{m=0}^{M-1} \bar{k}^2 p_{k,m}, \notag \\
    \mathcal{I}(\bm{\theta})_{1, 2} = \frac{8\pi^2 \mathrm{SNR}_s}{KM} \sum_{k=0}^{K-1} \sum_{m=0}^{M-1} \bar{k} \bar{m} p_{k,m}, \notag \\
    \mathcal{I}(\bm{\theta})_{2, 2} = \frac{8\pi^2 \mathrm{SNR}_s}{M^2} \sum_{k=0}^{K-1} \sum_{m=0}^{M-1} \bar{m}^2 p_{k,m},
    \label{eq:fisher_element}
\end{align}
where $p_{k,m} = |X[k, m]|^2$, and we note that $\mathcal{I}(\bm{\theta})_{1, 2} = \mathcal{I}(\bm{\theta})_{2, 1}$ holds. 
The CRLB values of the delay and Doppler estimates are given by the diagonal elements of the inverse of the Fisher information matrix, $\mathcal{I}^{-1}(\bm{\theta})$, which can be expressed as:
\begin{equation}
    \text{CRLB}(n^*) = \frac{\mathcal{I}(\bm{\theta})_{1, 1}}{\text{det}(\mathcal{I}(\bm{\theta}))}, \quad \text{CRLB}(v^*) = \frac{\mathcal{I}(\bm{\theta})_{2, 2}}{\text{det}(\mathcal{I}(\bm{\theta}))},
    \label{eq:crlb_det}
\end{equation}
where $\text{det}(\cdot)$ denotes the determinant of the matrix. By substituting \eqref{eq:fisher_element} into \eqref{eq:crlb_det}, we obtain the final expression in \eqref{eq:crlb_express} which completes the derivation.

\end{document}